\documentclass[fleqn,11pt]{article}
\usepackage{amscd,amsmath,amssymb,verbatim}
\usepackage{amsthm}
\usepackage{amssymb}
\usepackage{pifont}
\usepackage[pdftex]{graphicx}
\usepackage{epstopdf}
\usepackage[T1]{fontenc}
\usepackage{ae,aecompl}
\usepackage{makecell}
\usepackage{subcaption}
\usepackage{array}
\usepackage{mathrsfs}
\usepackage[nohead]{geometry}
\usepackage[singlespacing]{setspace}
\usepackage{rotating}
\usepackage{multirow}
\usepackage{threeparttable}
\usepackage{enumerate}
\usepackage{placeins}
\usepackage{color, colortbl}
\usepackage{accents}
\usepackage{pdflscape}
\usepackage{longtable}
\usepackage{natbib}
\usepackage{graphicx}
\usepackage{subcaption}
\usepackage{caption}
\usepackage{adjustbox}   
\usepackage{booktabs}    
\usepackage{caption}     

\usepackage{tabularx}
\usepackage{array}
\usepackage{booktabs}
\usepackage{float}

\usepackage{subcaption}
\usepackage{float}
\numberwithin{equation}{section}

\usepackage{hyperref}

\makeatletter
\newcommand{\distas}[1]{\mathbin{\overset{#1}{\kern\z@\sim}}}%
\newsavebox{\mybox}\newsavebox{\mysim}
\newcommand{\distras}[1]{%
  \savebox{\mybox}{\hbox{\kern3pt$\scriptstyle#1$\kern3pt}}%
  \savebox{\mysim}{\hbox{$\sim$}}%
  \mathbin{\overset{#1}{\kern\z@\resizebox{\wd\mybox}{\ht\mysim}{$\sim$}}}%
}
\makeatother

\renewcommand{\hat}[1]{\widehat{\text{$#1$}}}
\makeatletter
\newsavebox\myboxA
\newsavebox\myboxB
\newlength\mylenA

\renewcommand*\bar[2][0.85]{%
    \sbox{\myboxA}{$\m@th#2$}%
    \setbox\myboxB\null
    \ht\myboxB=\ht\myboxA%
    \dp\myboxB=\dp\myboxA%
    \wd\myboxB=#1\wd\myboxA
    \sbox\myboxB{$\m@th\overline{\copy\myboxB}$}
    \setlength\mylenA{\the\wd\myboxA}
    \addtolength\mylenA{-\the\wd\myboxB}%
    \ifdim\wd\myboxB<\wd\myboxA%
       \rlap{\hskip 0.5\mylenA\usebox\myboxB}{\usebox\myboxA}%
    \else
        \hskip -0.5\mylenA\rlap{\usebox\myboxA}{\hskip 0.5\mylenA\usebox\myboxB}%
    \fi}
\makeatother
\makeatletter
\def\@biblabel#1{\hspace*{-\labelsep}}
\makeatother
\hoffset 
\voffset
\author{Ariane N. Meli Chrisko\thanks{University of Glasgow, UK, email: a.meli-chrisko.1@research.gla.ac.uk}  \and Philipp Otto\thanks{University of Glasgow, UK, email: philipp.otto@glasgow.ac.uk}  }\medskip

\date{\today}
\title{Spatiotemporal dynamics of wind-speed volatility}

\begin{document}
\maketitle
\sloppy

\singlespacing

\begin{abstract}
\noindent
Wind-speed time series exhibit substantial temporal variability and spatial dependence, yet volatility dynamics across monitoring networks remain relatively unexplored. This study investigates the spatiotemporal behaviour of wind-speed volatility using daily observations from a dense network of 141 stations in Northern Italy over the period 2016--2021. Wind speeds are observed at two heights (10\,m and 100\,m), allowing for the analysis of both spatial interactions and vertical dependence.
To capture these features, we adopt a parsimonious spatiotemporal GARCH-type framework in which conditional variance depends on past local shocks and spatially aggregated information from neighbouring stations. The approach combines a spatial mean specification with structured volatility models based on alternative weight matrices reflecting geographical proximity and atmospheric flow direction.
Empirical results show that properly accounting for spatial dependence in the mean equation substantially improves residual diagnostics. The forecasting analysis further reveals that the relative performance of spatial volatility specifications is largely driven by the mean model: flexible local structures perform better when residual spatial dependence remains, whereas parsimonious distance-based specifications provide the most reliable out-of-sample forecasts once spatial interactions are adequately captured.
Volatility persistence is found to increase with height, indicating more stable dynamics at turbine hub levels. A multivariate extension provides additional evidence of cross-height dependence, with asymmetric interactions in the mean and more balanced spillovers in the volatility dynamics.
These findings highlight the importance of jointly modelling spatial and vertical dependence, and show that reliable volatility forecasts can be achieved through a combination of well-specified mean dynamics and parsimonious spatial structures.
\end{abstract}



\onehalfspacing

\section{Introduction}

Wind speed exhibits substantial variability across both time and space, and this variability plays a central role in wind-power generation, grid integration, and electricity market participation. In many applications, uncertainty in wind conditions is as consequential as the expected level itself. Because wind power is produced at geographically fixed sites and traded in markets where revenues and imbalance penalties depend on forecast errors, reliable location-specific measures of variability are essential for managing both operational and financial risk. Accurate modelling of wind-speed volatility is therefore a key component of short-term decision-making.

Beyond forecasting uncertainty, modelling wind-speed volatility has several important practical applications. Knowledge of the first and second moment dynamics of wind speed can inform the design of wind-resistant structures and improve engineering assessments of wind loads. In the context of wind-energy production, volatility estimates can support operational decisions such as turbine cut-in and cut-out strategies, maintenance scheduling, and production planning. Understanding wind variability can also guide turbine placement, measurement-network design, and site selection for wind farms. Moreover, volatility information can be used to evaluate the probability distribution of wind-power output and turbine operating conditions, which is essential for reliability assessment and electricity production planning \citep{ewing2006time, liu2013empirical}.
While a large fraction of the wind-energy literature has focused on forecasting the conditional mean of wind speed or power, increasing attention has been paid to modelling its conditional variance, or volatility. Time-varying volatility directly affects the width and calibration of prediction intervals and is closely linked to ramp events, turbulence intensity, mechanical loads on turbines, and uncertainty in wind-power output. Because wind is a spatially continuous atmospheric process driven by large-scale circulation and local terrain effects, variability at neighbouring sites is inherently correlated. This implies that uncertainty cannot be adequately captured by models that treat locations independently, motivating the use of statistical frameworks that jointly account for temporal volatility dynamics and spatial dependence.

Early empirical studies reported that wind-speed residuals exhibit volatility clustering, a characteristic feature of autoregressive conditional heteroskedasticity. Applications of ARMA--GARCH-type models have shown that conditional variance in wind-speed and wind-power series is time-varying and can be effectively captured using GARCH-type specifications \citep{lojowska2010advantages, liu2013empirical}. For instance, \cite{erdem2011arma} demonstrated that incorporating conditional variance dynamics improves the statistical representation of wind-speed series, while \cite{cripps2003bivariate} modelled horizontal wind components using a bivariate GARCH framework and showed that volatility modelling enhances the representation of high-frequency fluctuations. Extending this line of research, \cite{jeontaylor2012} employed VARMA--GARCH models to jointly model wind speed and wind direction, reporting improvements in density forecasting performance.
A growing body of work has also investigated asymmetric volatility effects in wind processes, motivated by the observation that abrupt changes in wind speed, particularly sharp decreases, may trigger stronger volatility responses. To capture this behaviour, several studies have considered asymmetric GARCH variants such as EGARCH, TGARCH, and APARCH. Empirical applications to wind-speed and wind-power data report that allowing for asymmetric responses to shocks can improve the representation of volatility bursts and turbulence episodes, particularly in probabilistic forecasting settings \citep{ewing2006time,ziel2016forecasting,shen2016wind,chen2019applied}.

Despite the availability of increasingly sophisticated GARCH variants, the literature consistently reports that low-order specifications such as GARCH(1,1) and EGARCH(1,1) capture the dominant features of wind-speed volatility. This finding mirrors well-known results from financial econometrics: \citet{bollerslev1986garch} showed that GARCH(1,1) provides an effective approximation for a wide range of heteroskedastic processes, while \citet{hansenlunde2005} demonstrated that more complex volatility models rarely deliver systematic improvements in forecasting accuracy. Similar conclusions have been reported in meteorological and energy-related applications, where parsimonious GARCH-type models offer a favourable balance between flexibility, numerical stability, and interpretability \citep{erdem2011arma,liu2011comprehensive,carapella2016meteo}.
More recent contributions have combined GARCH-type models with alternative methodologies, including wavelet decompositions, neural networks, regime-switching dynamics, and long-memory structures \citep{cao2012forecasting,ailliot2015non,pradhan2020wind,yilmaz2024long,band2025two}. Although such hybrid approaches may deliver improvements for specific forecasting horizons or locations, they are often computationally demanding and less transparent, particularly in large-scale or spatially distributed settings.

Importantly, the vast majority of existing studies estimate volatility models univariately, treating each location independently. This modelling strategy ignores spatial dependence in conditional variance, an assumption that is difficult to reconcile with the physical nature of wind as a spatially continuous atmospheric process driven by large-scale circulation patterns and terrain effects. While spatial statistical methods have been widely used to model the conditional mean through interpolation, kriging, or hierarchical approaches, explicit modelling of spatiotemporal dependence in volatility remains relatively limited \citep{gneiting2007geostatistical,obakrim2022statistical,rychlik2019spatio}.
This gap motivates the present study. We first analyse temporal volatility dynamics at individual stations using univariate GARCH and EGARCH models, assessing volatility persistence and asymmetry across measurement heights. Building on these results, we then apply a spatiotemporal volatility framework that explicitly accounts for spatial spillovers through weight matrices reflecting geographical proximity and atmospheric transport mechanisms. Finally, we extend this framework to a multivariate setting across measurement heights, allowing for joint modelling of horizontal spatial interactions and vertical dependence between 10\,m and 100\,m wind speeds.

The contribution of this paper is threefold. First, we provide a systematic empirical investigation of spatiotemporal volatility dynamics in wind-speed data across a dense monitoring network. Second, we show that the relative performance of spatiotemporal volatility models is closely linked to the specification of the mean equation: flexible spatial structures appear advantageous when residual spatial dependence remains, whereas parsimonious distance-based specifications provide more reliable forecasts once spatial interactions are properly accounted for. Third, we demonstrate that cross-height dependence plays a non-negligible role, with asymmetric interactions in the mean and more stable spillovers in the volatility dynamics.
The remainder of the paper is organised as follows. Section~\ref{wind_data_source} describes the wind-speed dataset and its spatial coverage. Section~\ref{sec:univariate_garch} presents the station-wise univariate volatility analysis. Section~\ref{stgarch_sec} and Section~\ref{sec5:multivariatelogarch} present the spatiotemporal and multivariate volatility analyses, respectively, and discuss the empirical results. Section~\ref{sec:conclusion} concludes.

\section{Wind Speed Data}\label{wind_data_source}

\begin{figure}[t]
\centering
\includegraphics[width=0.95\textwidth]{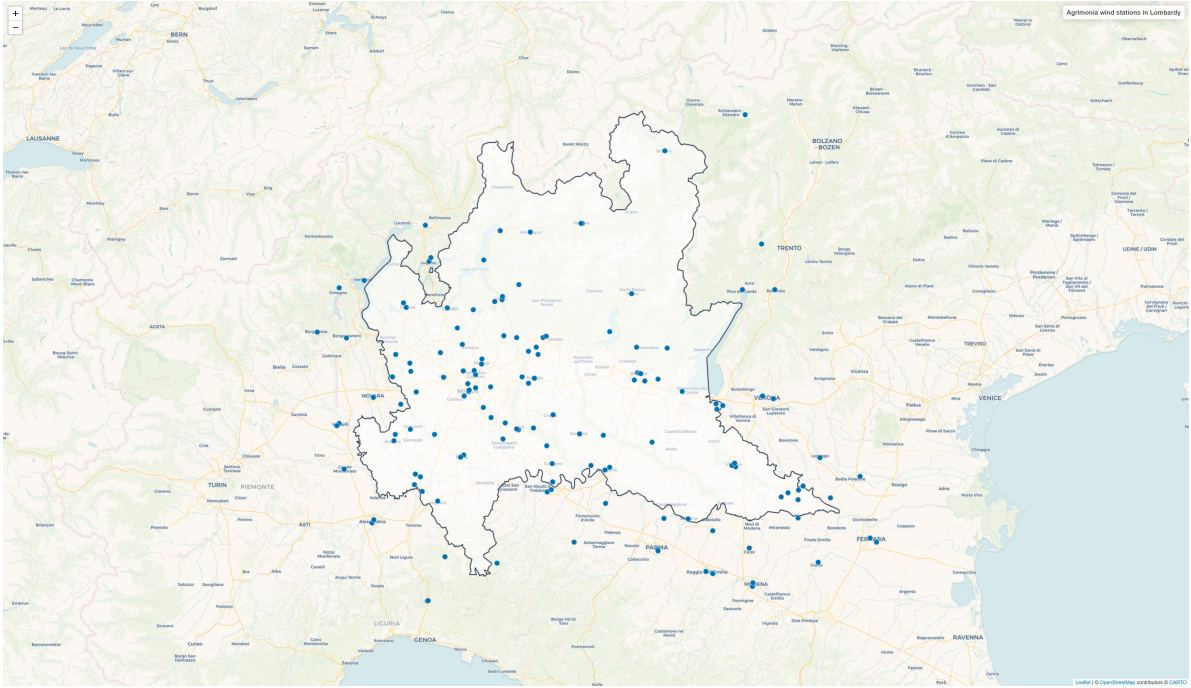}
\caption{Spatial distribution of Agrimonia wind stations in Lombardy.}
\label{wind_stations}
\end{figure}

The wind speed data used in this study come from the
\emph{Agrimonia Dataset}~\cite{fasso2023agrimonia}, a high-resolution spatiotemporal database integrating air quality, meteorological, emission, livestock, and land-use information over Northern Italy. The dataset covers the Lombardy region and its surrounding area, defined by applying a $0.3^{\circ}$ spatial buffer around the regional boundaries, and spans the period from January~2016 to December~2021. Meteorological variables are derived from a combination of ground-based observations and reanalysis products and are provided at a daily frequency.
Wind speed is available at multiple heights and in different aggregated forms. In this study, we focus on daily mean horizontal wind speed measured at heights of 10~m and 100~m above the earth \footnote{In this study, we choose $\textbf{ws10}$ and $\textbf{ws100}$ to refer to wind-speed at 10~m and 100~m  respectively}. The 10~m wind speed corresponds to the standard meteorological reference height, while the 100~m wind speed is representative of typical wind-turbine hub heights and is therefore particularly relevant for wind-energy applications.
Let $u_{t}(s_i,\tau)$ denote the instantaneous wind speed observed at station $s_i$ and intra-day time index $\tau$ on day $t$. The daily mean wind speed is defined as
\begin{equation}
\label{eq:daily_mean_wind}
Y_{t}(s_i) = \frac{1}{M_t} \sum_{\tau=1}^{M_t} u_{t}(s_i,\tau),
\end{equation}
where $M_t$ denotes the number of observations available on day $t$. This aggregation suppresses very high-frequency fluctuations while
preserving large-scale atmospheric variability, making the resulting series
well-suited for time-series volatility modelling. To avoid confounding atmospheric variability with extreme gust behaviour, daily maximum wind speed variables are not considered in the present analysis.
The dataset contains observations from $N = 141$ monitoring stations distributed across Lombardy and neighbouring regions (see Figure~\ref{wind_stations}). Station locations are irregularly spaced, reflecting the operational monitoring network. For each station, observations are available from 2016-01-01 to 2021-12-31, resulting in 2192 daily observations per station. Missing values in the meteorological series are limited and have been addressed in the Agrimonia preprocessing pipeline using interpolation and imputation procedures described
in~\cite{fasso2023agrimonia} and data can be found here \url{https://zenodo.org/records/7956006}.
\begin{table}[t]
\centering
\caption{Descriptive statistics of daily mean wind speed (2016--2021).}
\label{tab:wind_descriptive}
\begin{tabular}{lcccccccc}
\hline
Height & $T$ & N& Median & Mean & IQR & SD  & min & max\\
\hline
10 m  & 2192 & 141 &1.313 & 1.475 & 0.717 & 0.689 & 0.2309 & 7.036  \\
100 m & 2192 & 141 & 2.245 & 2.584 & 1.484 & 1.318 & 0.4357 & 13.050 \\
\hline
\end{tabular}
\end{table}

Table~\ref{tab:wind_descriptive} summarises the distribution of daily mean wind speed over the full sample period. Wind speeds at 100~m are consistently higher and more variable than those observed at 10~m, reflecting reduced surface friction and stronger atmospheric flow at higher elevations. At both heights, the mean exceeds the median, indicating moderate right-skewness in the marginal distributions. Moreover, the substantially larger interquartile range and standard deviation at 100~m highlight the increased variability of wind conditions at turbine hub heights, motivating separate modelling of volatility dynamics across measurement heights.
While these descriptive statistics provide a global overview of wind speed
behaviour across the study region, they do not capture the pronounced temporal variability observed within individual station series. In particular, wind speed exhibits periods of heightened and subdued variability over time, a feature that cannot be adequately described by constant-variance models. To characterise these dynamics, we begin by analysing the temporal volatility structure of wind speed at individual stations using univariate GARCH-type models. This step provides a benchmark for understanding time-varying heteroskedasticity before introducing spatial dependence and spatiotemporal
extensions. 

\section{Station-wise univariate GARCH and EGARCH analysis}
\label{sec:univariate_garch}
Let $Y_t(s_i)$ denote the observed daily mean wind speed at time $t$ at station
$s_i$, $i = 1,\ldots,N$. Wind-speed series typically exhibit pronounced
deterministic seasonality and short-term serial dependence. To isolate the
stochastic component relevant for volatility modelling, the conditional mean is removed using a two-step procedure applied uniformly across all stations.

In the first step, the seasonal component is extracted using Seasonal-Trend
decomposition based on LOESS (STL; \citealp{cleveland1990stl}). The STL
decomposition represents the series as

\begin{equation}
Y_t(s_i) = S_t(s_i) + T_t(s_i) + \tilde{Y}_t(s_i),
\end{equation}

where $S_t(s_i)$ denotes the seasonal component, $T_t(s_i)$ a smooth trend,
and $\tilde{Y}_t(s_i)$ the remainder.

In the second step, residual serial dependence in the deseasonalised series is
removed by fitting a first-order autoregressive model AR(1),

\begin{equation}
\tilde{Y}_t(s_i) = \phi_i \tilde{Y}_{t-1}(s_i) + e_t(s_i),
\qquad |\phi_i| < 1,
\end{equation}

where $e_t(s_i)$ is a zero-mean innovation process. The residuals $e_t(s_i)$
represent the stochastic component of wind-speed variability and are used as
inputs for volatility modelling.

\begin{figure}[h!]
  \centering
  \includegraphics[width=0.95\linewidth]{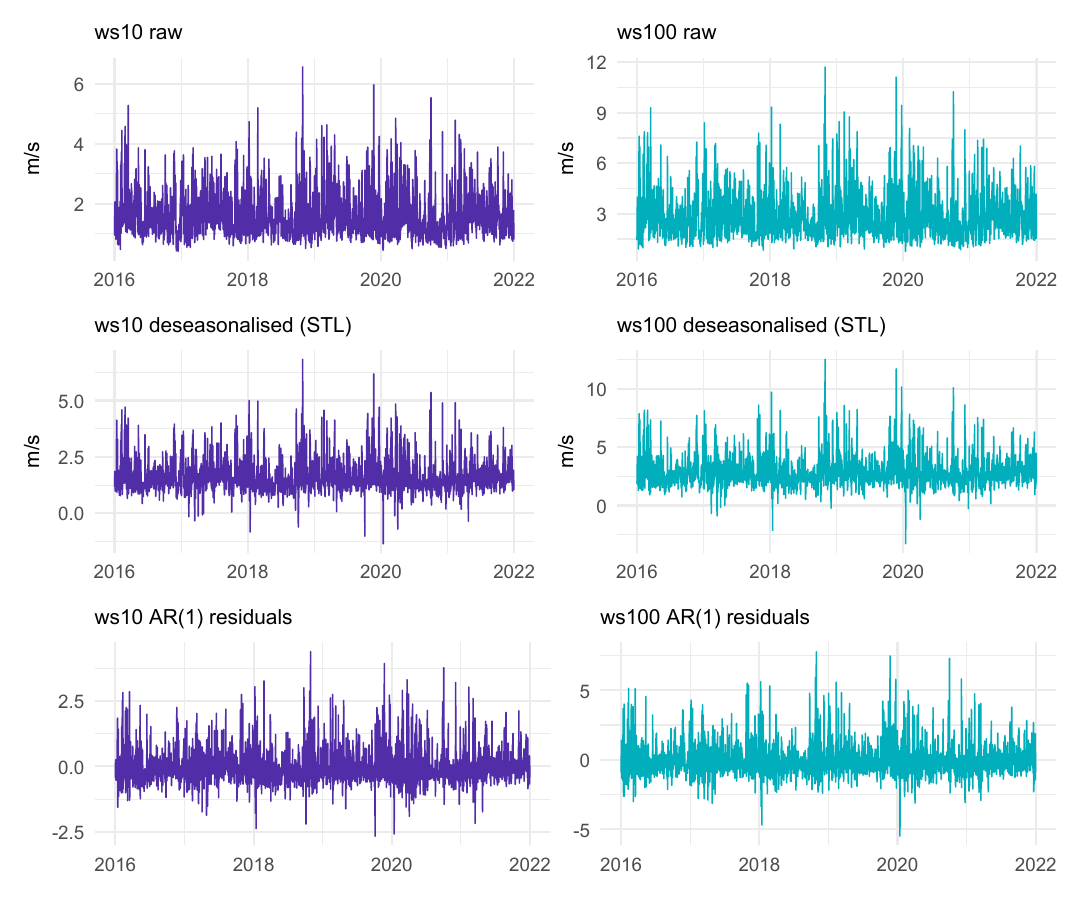}
  \caption{Raw wind speed, STL-deseasonalised series, and AR(1) residuals at 10\,m and 100\,m for a representative station (Station~1264).}
  \label{fig:stl_ar1_pipeline}
\end{figure}

\begin{figure}[h!]
  \centering
  \includegraphics[width=0.95\linewidth]{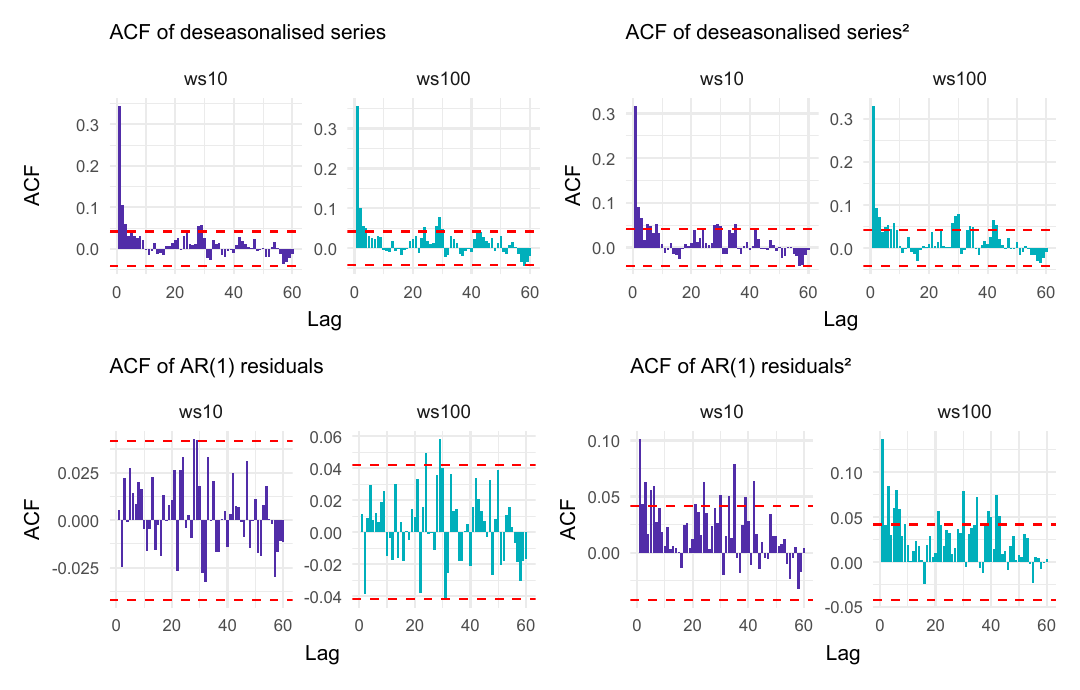}
  \caption{Autocorrelation functions (ACF) and squared ACF of deseasonalised wind speed and AR(1) residuals at 10\,m and 100\,m (Station~1264).}
  \label{fig:acf_diagnostics}
\end{figure}

Figures~\ref{fig:stl_ar1_pipeline} and~\ref{fig:acf_diagnostics} illustrate the preprocessing steps and the temporal dependence structure of the wind-speed series for a representative monitoring station. Figure~\ref{fig:stl_ar1_pipeline} displays the raw daily wind-speed series together with the STL-deseasonalised series and the residuals obtained from an AR(1) model. The STL decomposition removes low-frequency seasonal components while preserving short-term fluctuations in wind speed. After deseasonalisation and AR(1) filtering, the residual series oscillate around zero and exhibit no obvious deterministic structure, indicating that the main components of the conditional mean have been adequately removed.

Figure~\ref{fig:acf_diagnostics} reports the sample autocorrelation functions (ACF) of the deseasonalised series and the AR(1) residuals, together with the corresponding ACF of the squared series. The deseasonalised wind-speed series display clear temporal dependence, as evidenced by significant autocorrelation at short lags. In contrast, the AR(1) residuals exhibit substantially reduced linear autocorrelation, suggesting that the simple autoregressive specification captures a large portion of the mean dynamics. However, the autocorrelation functions of the squared residuals remain statistically significant at several lags, revealing the presence of conditional heteroskedasticity. This pattern, characterised by weak autocorrelation in the residuals but persistent dependence in their squared values, is a typical empirical signature of volatility clustering and provides empirical support for the use of GARCH-type models to describe wind-speed variability.

Station-level ARCH diagnostics based on the residuals $e_t(s_i)$ using the
ARCH--LM test provide strong evidence of time-varying volatility in
wind-speed dynamics. For wind speed measured at 10\,m height,
approximately $91\%$ of stations reject the null hypothesis of
homoskedasticity, indicating widespread volatility clustering.
At 100\,m height, the null is rejected for all stations,
suggesting that conditional heteroskedasticity is pervasive aloft.
Additionally, all series exhibit kurtosis greater than three,
indicating heavy-tailed distributions and a relatively high
frequency of extreme wind events.

\medskip

Volatility dynamics on $e_t(s_i)$ are first modelled using the standard
GARCH(1,1) specification introduced by \citet{bollerslev1986garch},
\begin{align}
e_t(s_i) &= \sqrt{h_t(s_i)}\, z_t(s_i), \qquad z_t(s_i) \sim \mathcal{N}(0,1), \\
h_t(s_i) &= \omega + \alpha e_{t-1}^2(s_i) + \beta h_{t-1}(s_i),
\end{align}
where $h_t(s_i)$ denotes the conditional variance and $\omega > 0$,
$\alpha \ge 0$, and $\beta \ge 0$. This formulation implies a symmetric
response of volatility to past shocks, as only squared innovations enter the
variance equation. To accommodate potential asymmetry in volatility responses, we also consider
the exponential GARCH (EGARCH) model proposed by \citet{nelson1991egarch}. In the EGARCH(1,1) specification, the logarithm of the conditional variance evolves as
\begin{equation}
\log h_t(s_i)
=
\omega
+ \beta \log h_{t-1}(s_i)
+ \alpha \left( |z_{t-1}(s_i)| - \mathbb{E}|z_{t-1}(s_i)| \right)
+ \gamma z_{t-1}(s_i).
\end{equation}
The parameter $\gamma$ captures asymmetric volatility effects, allowing
positive and negative shocks of equal magnitude to have different impacts on
future variance. Unlike the standard GARCH model, EGARCH does not require
positivity constraints on the parameters.

\begin{figure}[h!]
\centering

\begin{subfigure}{0.48\textwidth}
  \centering
  \includegraphics[width=\linewidth]{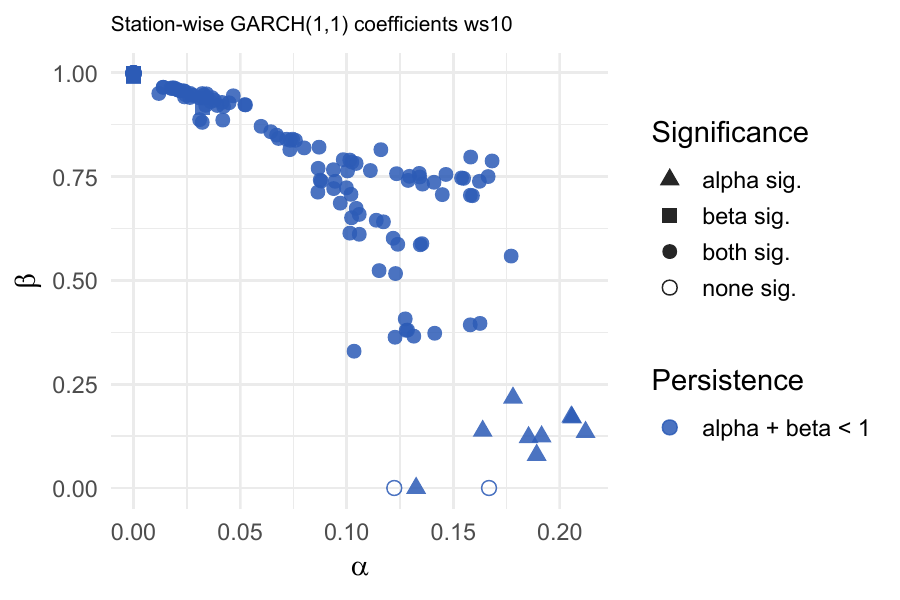}
  \caption{GARCH(1,1), 10\,m}
\end{subfigure}
\hfill
\begin{subfigure}{0.48\textwidth}
  \centering
  \includegraphics[width=\linewidth]{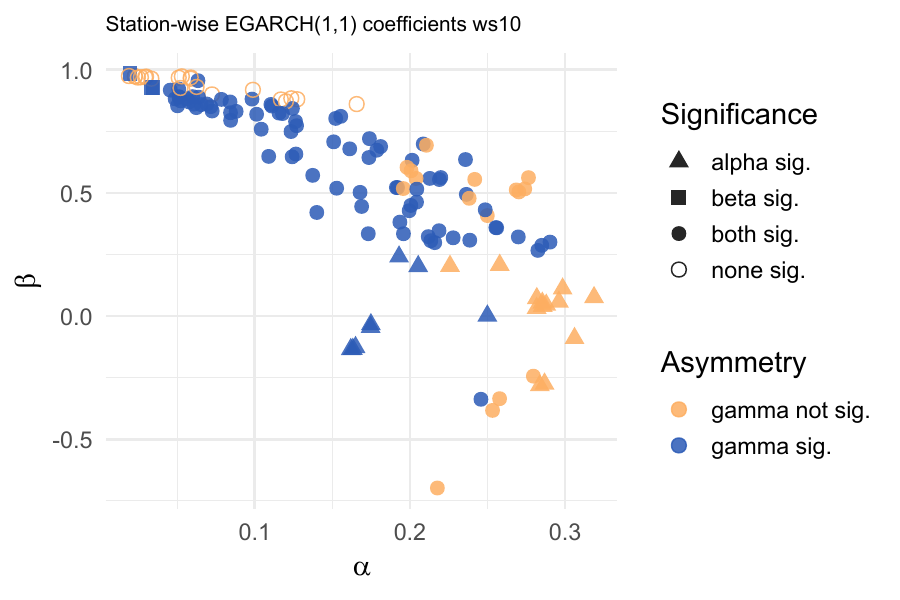}
  \caption{EGARCH(1,1), 10\,m}
\end{subfigure}

\vspace{0.3cm}

\begin{subfigure}{0.48\textwidth}
  \centering
  \includegraphics[width=\linewidth]{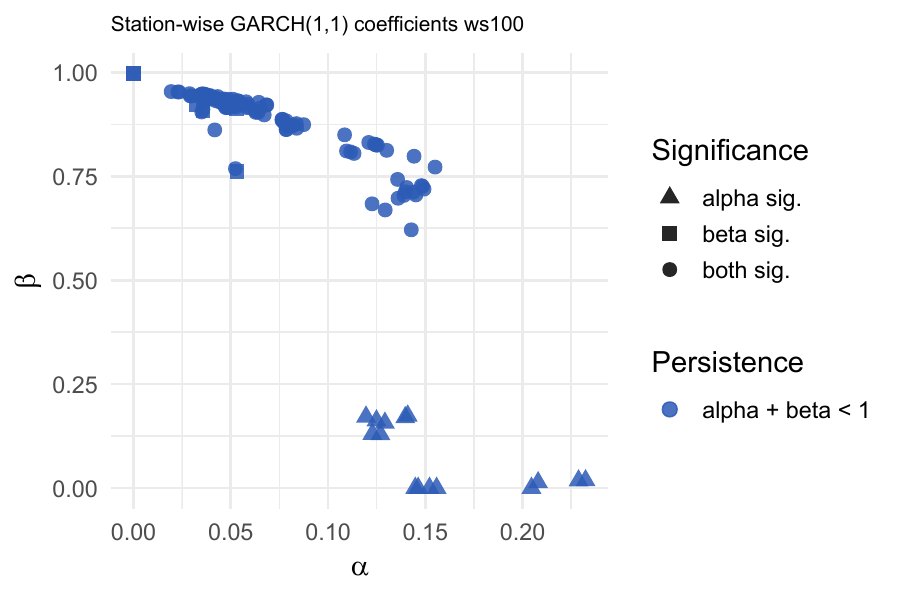}
  \caption{GARCH(1,1), 100\,m}
\end{subfigure}
\hfill
\begin{subfigure}{0.48\textwidth}
  \centering
  \includegraphics[width=\linewidth]{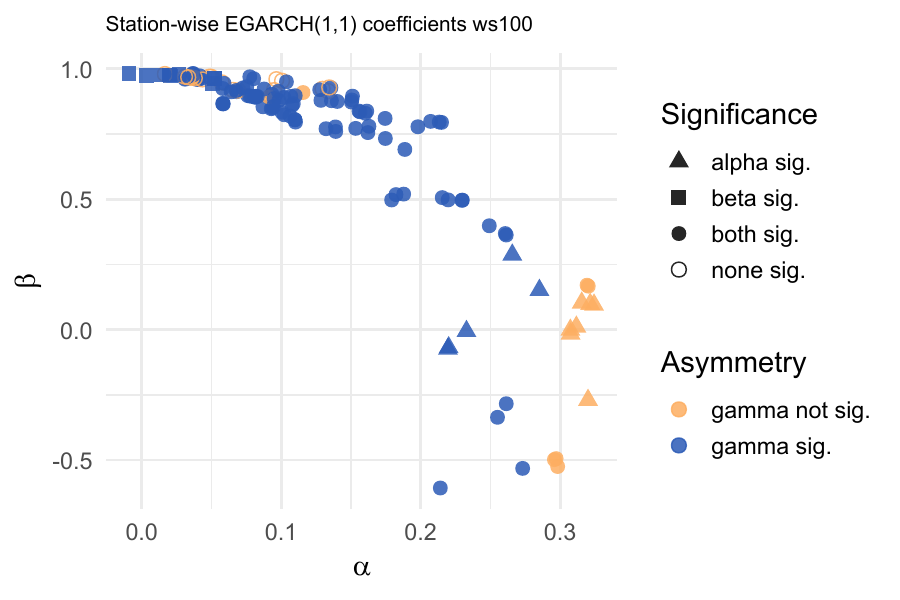}
  \caption{EGARCH(1,1), 100\,m}
\end{subfigure}
\caption{
Station-wise GARCH(1,1) and EGARCH(1,1) coefficient estimates for wind speed at 10\,m (top row) and 100\,m (bottom row). Points are shown in the $(\hat{\alpha},\hat{\beta})$
plane. Shapes indicate statistical significance of the ARCH and GARCH
parameters. Colours indicate volatility persistence in the GARCH specification or statistical significance of the asymmetry parameter $\gamma$ in the EGARCH model.
}
\label{fig:garch_egarch_stationwise}
\end{figure}

Figure~\ref{fig:garch_egarch_stationwise} illustrates station-wise GARCH(1,1) and
EGARCH(1,1) parameter estimates for wind speed at 10\,m and 100\,m, fitted on the
training sample of 1827 observations (2016--2020). 
At both heights, the estimated coefficients reveal substantial cross-sectional
heterogeneity in volatility dynamics across monitoring stations. Most stations
exhibit large values of the persistence parameter $\beta$, often combined with
moderate ARCH effects, indicating a high degree of volatility persistence in
wind-speed dynamics. A comparison across heights suggests that volatility persistence tends to be
stronger at 100\,m than at 10\,m, consistent with smoother atmospheric dynamics
away from the surface layer where turbulence and frictional effects dominate.
The EGARCH specification further reveals that asymmetric volatility responses
are statistically significant for a large fraction of stations, particularly at
10\,m height, indicating that positive and negative wind-speed shocks may have
different impacts on future volatility.

\begin{table}[!ht]
\centering
\caption{Training-sample information-criteria comparison for ws10 and ws100.
Percentages indicate the share of stations for which EGARCH(1,1) is preferred
over GARCH(1,1).}
\label{tab:ic_ws}
\begin{tabular}{lrr}
\toprule
Height & EGARCH pref. (AIC) [\%] & EGARCH pref. (BIC) [\%] \\
\midrule
ws10  & 96.45 & 87.23  \\
ws100 & 91.48 & 76.59  \\
\bottomrule
\end{tabular}
\end{table}

\begin{table}[!ht]
\centering
\caption{Training sample Ljung--Box diagnostic pass rates for ws10 and ws100
(5\% level). Tests are applied to standardised residuals and squared standardised residuals.
Values are the percentages of stations with $p$-values greater than 0.05.}
\label{tab:lb_ws}
\begin{tabular}{llrrrr}
\toprule
Height & Model & ($\hat{e}_t$ Lag 10) [\%] & ($\hat{e}_t$ Lag 20) [\%] & ($\hat{e}_t^2$ Lag 10) [\%] & ($\hat{e}_t^2$ Lag 20) [\%] \\
\midrule
ws10  & GARCH & 81.56 & 97.87 & 86.52 & 93.61  \\
ws10  & EGARCH & 77.30 & 94.32 & 96.45 & 99.29 \\
\midrule
ws100 & GARCH & 77.30 & 89.36 & 86.52 & 91.48 \\
ws100 & EGARCH & 71.63 & 84.39 & 84.39 & 92.90 \\
\bottomrule
\end{tabular}
\end{table}

Table~\ref{tab:ic_ws} confirms that EGARCH is preferred over GARCH for the
majority of stations at both heights, with the preference being stronger at
10\,m than at 100\,m. Specifically, EGARCH is selected for 96.45\% of stations
according to AIC and 87.23\% according to BIC at 10\,m, compared with
91.48\% and 76.59\%, respectively, at 100\,m. This indicates that allowing
for asymmetric volatility responses improves the in-sample fit, particularly
near the surface.
The Ljung--Box diagnostic pass rates reported in Table~\ref{tab:lb_ws}
provide a complementary perspective. For both heights, the GARCH model
achieves slightly higher pass rates on the standardised residuals,
indicating a more effective removal of serial dependence in the first
moment. In contrast, EGARCH generally yields higher pass rates for
squared residuals, especially at 10\,m, suggesting that it captures
conditional heteroskedasticity more effectively.
Taken together, these results indicate that wind-speed volatility is
persistent and heterogeneous across monitoring stations, with asymmetric
effects playing a non-negligible role at both heights, although more
pronounced near the surface. The station-wise models, therefore, provide
useful benchmark evidence on the temporal dynamics of wind-speed
volatility.

However, these models are estimated independently for each station and
therefore ignore potential spatial interactions across locations.
Wind fields evolve through atmospheric transport processes and are
strongly influenced by local terrain and boundary-layer dynamics,
implying that neighbouring stations may experience correlated shocks
and similar volatility patterns.
Consequently, modelling wind-speed variability using purely
station-wise specifications may overlook important cross-location
dependence. A modelling framework that explicitly incorporates
spatial interactions is therefore required to capture the joint
evolution of wind-speed variability across the monitoring stations.
In the following section, we therefore investigate spatial dependence in the data and introduce a spatiotemporal volatility framework that allows both temporal persistence and spatial spillovers in wind-speed volatility.

\section{On a spatiotemporal analysis of wind speed volatility} \label{stgarch_sec} 

The previous station-wise analysis provides evidence of persistent and heterogeneous volatility dynamics in wind-speed series. While such univariate models are useful for characterising temporal behaviour at individual monitoring stations, they do not account for spatial interactions that naturally arise in atmospheric processes. Wind-speed dynamics are inherently spatial: shocks generated by local gusts, pressure gradients, or moving weather systems can propagate across neighbouring locations, leading to spatially correlated patterns in both the mean and variance.
To incorporate these interactions, we adopt a parsimonious spatiotemporal specification in which the dynamics at each location depend on both past local behaviour and spatially aggregated information from neighbouring stations. Specifically, we implement the STARMAGARCH framework \citet{Holleland2020stationary}, which combines a Spatiotemporal Autoregressive Moving Average (STARMA) mean equation with a Spatiotemporal GARCH-type volatility process. This framework provides a flexible yet computationally tractable way to introduce structured spatial dependence without resorting to fully multivariate volatility models.
In line with the empirical implementation adopted here, we restrict attention to a first-order spatial autoregressive and moving-average structure in the conditional mean together with a first-order spatiotemporal GARCH recursion for the conditional variance, resulting in a STARMAGARCH$(1,1,1,1)$ specification. Previous comparative work \cite{chrisko2026comparative} indicates that such structured spatiotemporal formulations can achieve competitive predictive performance while remaining substantially more parsimonious than heavily parameterised multivariate alternatives such as the DCC-GARCH.

Let $\mathbf{e}_t = (e_t(s_1),\ldots,e_t(s_N))^\top$ denote the $N\times1$ vector of residuals obtained from the preprocessing procedure described in Section~\ref{sec:univariate_garch}. Let us note that $\mathbf{e}_t$ could be the residuals from any appropriate mean model.  Furthermore, let $\mathbf{h}_t = (h_t(s_1),\ldots,h_t(s_N))^\top$ denote the vector of conditional variances across the monitoring stations at time $t$. The joint spatiotemporal model is specified as 
\begin{align} \mathbf{e}_t - \boldsymbol{\mu} &= \phi\,\mathbf{W}(\mathbf{e}_{t-1}-\boldsymbol{\mu}) + \theta\,\mathbf{W}\boldsymbol{\varepsilon}_{t-1} + \boldsymbol{\varepsilon}_t, \label{eq:mean_eq} \\ \boldsymbol{\varepsilon}_t &= \mathbf{h}_t^{1/2} \odot \boldsymbol{Z}_t, \qquad \boldsymbol{Z}_t \sim \text{i.i.d. }(\mathbf{0},\mathbf{I}), \label{eq:innov_eq} \\ \mathbf{h}_t &= \boldsymbol{\omega} + \alpha\,\mathbf{W}(\boldsymbol{\varepsilon}_{t-1} \odot \boldsymbol{\varepsilon}_{t-1}) + \beta\,\mathbf{W}\mathbf{h}_{t-1}, \label{eq:vol_eq} 
\end{align} 
where $\mathbf{W}$ is a row-standardised spatial weight matrix encoding the connectivity structure of spatial locations, $\boldsymbol{\omega}$ is a vector of station-specific intercepts, and $\odot$ denotes the Hadamard (element-wise) product. The parameters satisfy $\omega_i>0$, $\alpha\ge0$, and $\beta\ge0$ for $i=1,\ldots,N$. The parameters have natural interpretations in the context of spatial wind dynamics:
\begin{itemize}
    \item Mean dynamics ($\phi$, $\theta$): Capture spatiotemporal dependence in the mean equation, allowing wind-speed levels to evolve over time and propagate across locations through the spatial structure defined by $\mathbf{W}$.

    \item Spatial ARCH ($\alpha$): Represents the spatiotemporal propagation of shocks across locations. High-intensity events at one station contribute to increased conditional variance at neighbouring sites through the spatial aggregation of past squared residuals.

    \item Spatiotemporal persistence ($\beta$): Captures the persistence of volatility over time and its propagation across locations, reflecting sustained periods of elevated variability such as multi-day weather systems moving through the region.
\end{itemize}

Although higher-order specifications can in principle be formulated, estimation and stability enforcement become increasingly challenging in high-dimensional settings ($N=141$). Empirical evidence further suggests that such extensions rarely lead to meaningful improvements in one-step-ahead forecasting performance once spatial dependence is explicitly modelled. Consequently, the first-order STARMAGARCH specification represents a natural compromise between geographic detail and computational tractability.

\subsection{Spatial weights and testing spatial autocorrelation}
\label{sec:spatial_weights_moran}

\begin{figure}[h!]
\centering
\caption{Annual mean wind speed per station (ws10 vs ws100).}
\label{fig:annual_mean_wind}
\begin{subfigure}{0.49\textwidth}
  \includegraphics[width=\linewidth]{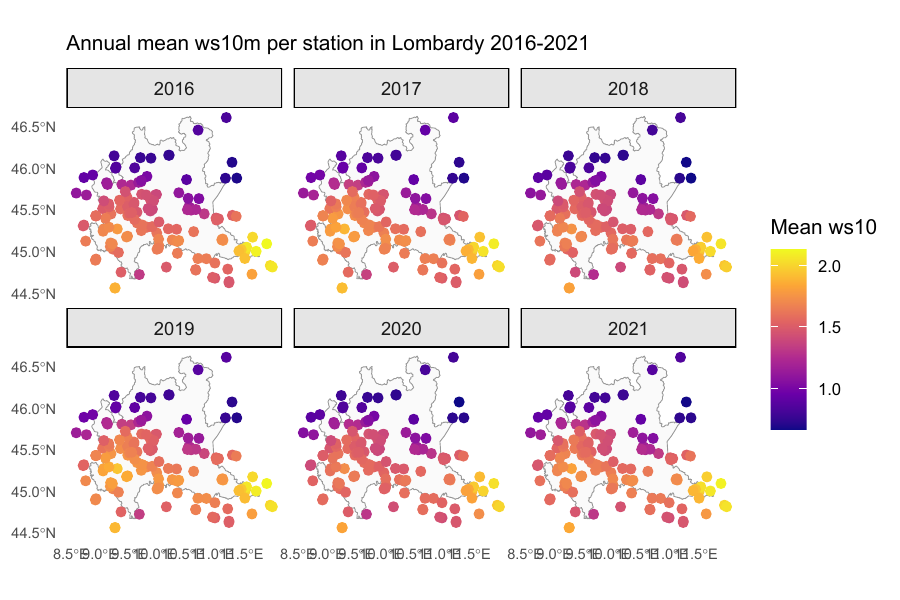}
  \caption{10\,m wind speed (ws10).}
\end{subfigure}\hfill
\begin{subfigure}{0.49\textwidth}
  \includegraphics[width=\linewidth]{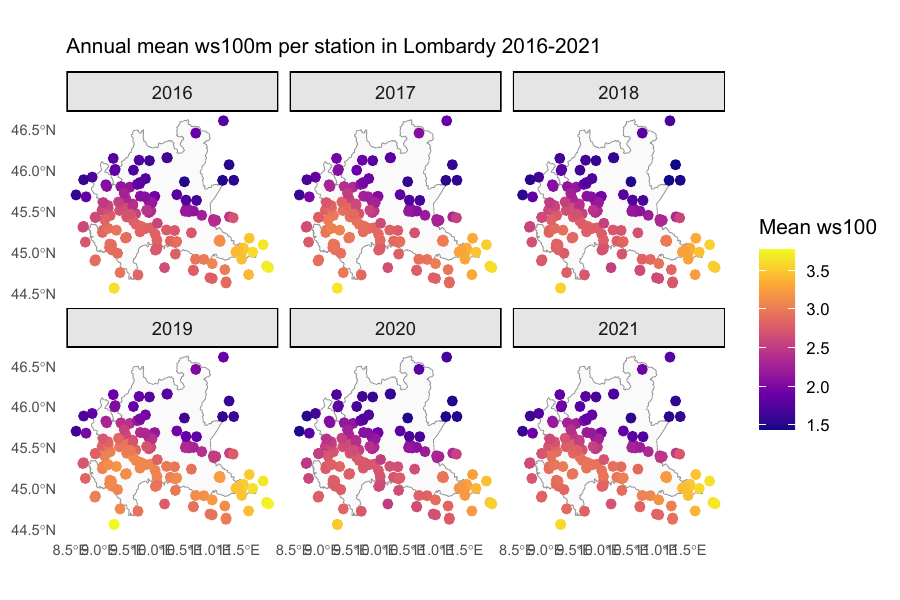}
  \caption{100\,m wind speed (ws100).}
\end{subfigure}
\end{figure}

Figure~\ref{fig:annual_mean_wind} displays the annual mean wind speed at each
monitoring station for the 10\,m and 100\,m measurement heights over the period
2016--2021. Even after aggregation over time, substantial spatial heterogeneity
is evident, with pronounced large-scale gradients across the study region.
In particular, stations located in the Alpine and pre-Alpine areas exhibit
systematically higher mean wind speeds than those situated in the Po Valley.
These spatial patterns persist across years and measurement heights,
suggesting that wind-speed variability is not spatially random.

While such maps are informative for visualising spatial structure, they do not
constitute formal evidence of spatial dependence. To quantify and test spatial
autocorrelation in wind-speed dynamics, we rely on Moran’s $I$ statistic,
which requires the specification of an appropriate spatial weight matrix.

Let $s_1,\ldots,s_N$ denote the spatial locations of the monitoring stations and
let $\mathbf{W} = (w_{ij})$ be a spatial weight matrix governing cross-sectional
dependence. All spatial weight matrices considered in this study are sparse,
time-invariant, and row-standardised such that $\sum_{j} w_{ij} = 1$ for all
$i$. Pairwise distances $d_{ij}$ are computed as Euclidean distances using
projected spatial coordinates.

\paragraph{(i) k-nearest-neighbour weights.}

For each location $s_i$, let $\mathcal{N}_k(i)$ denote the set of its $k$
geographically closest neighbours. The $k$-nearest-neighbour (k-NN) weight
matrix is defined as

\[
w_{ij}^{(\mathrm{kNN})}
=
\begin{cases}
1/k, & j \in \mathcal{N}_k(i), \\
0,   & \text{otherwise}.
\end{cases}
\]

This construction guarantees a fixed number of neighbours per location and
precludes isolated observations, making it particularly suitable for spatial
diagnostics such as Moran’s $I$
\citep{anselin1988spatial,bivand2008applied}.

\paragraph{(ii) Distance-band weights.}

Alternatively, spatial dependence is modelled using a distance-band structure,
where stations within a fixed radius $r$ are treated as neighbours:

\[
w_{ij}^{(\mathrm{band})}
=
\begin{cases}
1/n_i, & 0 < d_{ij} \le r, \\
0,     & \text{otherwise},
\end{cases}
\]

where $n_i$ denotes the number of neighbours of station $s_i$ within the
distance threshold $r$. This specification captures localised spatial
interactions within a finite geographical range and is commonly used in
regional and environmental studies \cite{anselin1988spatial}.

\paragraph{(iii) Directional (advection-based) weights.}

To account for directional propagation mechanisms inherent to atmospheric
processes, a directional spatial weight matrix is also considered.
Let $\theta_i$ denote the prevailing wind direction at station $s_i$.
Station $s_j$ is considered upwind of $s_i$ if the bearing from $s_i$
to $s_j$ lies within a cone of half-angle $\delta$ centred on $\theta_i$.

Directional weights are defined as

\[
w_{ij}^{(\mathrm{dir})}
=
\begin{cases}
\displaystyle
\frac{\exp(-d_{ij}/\phi)\cos(\Delta\theta_{ij})}
{\sum_{j' \in \mathcal{U}(i)} \exp(-d_{ij'}/\phi)\cos(\Delta\theta_{ij'})},
& j \in \mathcal{U}(i), \\
0, & \text{otherwise},
\end{cases}
\]

where $\Delta\theta_{ij}$ denotes the angular deviation from the prevailing wind
direction, $\phi>0$ controls the rate of spatial decay with distance,
and $\mathcal{U}(i)$ denotes the set of upwind neighbours of station $s_i$.
This construction captures anisotropic spatial dependence consistent with
atmospheric advection mechanisms \citep{wikle1999advection, merk2020estimation}.
Because directional weight matrices may be asymmetric and locally disconnected,
Moran’s $I$ is computed using row-standardised weights with appropriate
zero-neighbour handling.

Unlike spatial weight matrices commonly employed in financial applications,
which are often constructed from correlations or comovements of de-meaned
returns, the weight matrices considered here depend exclusively on the
geographical configuration of the monitoring network and prevailing
meteorological conditions. This reflects the physical nature of wind-speed
dynamics, where spatial dependence arises primarily from atmospheric transport
and terrain effects rather than from shared latent factors. Figure~\ref{fig:W_networks_row} in the Appendix presents spatial visualisations of the three spatial weight structures considered in this study.

\begin{table}[htbp]
\centering
\small
\caption{Moran’s $I$ tests for spatial autocorrelation in wind-speed residuals and volatility.}
\label{tab:moran_volatility_compact}
\begin{tabular}{l l c c c c}
\hline
 &  & \multicolumn{2}{c}{\textbf{ws10}} & \multicolumn{2}{c}{\textbf{ws100}} \\
\cline{3-6}
\textbf{Weight matrix} & \textbf{Proxy}
& \textbf{$I$} & \textbf{$Z$}
& \textbf{$I$} & \textbf{$Z$} \\
\hline
Distance-band (55 km)
& $\overline e$
& 0.404 & 17.29
& 0.454 & 19.43 \\
& $\overline{e^2}$
& 0.552 & 23.52
& 0.587 & 25.00 \\[0.3em]

$k$-nearest neighbours ($k=5$)
& $\overline e$
& 0.747 & 16.23
& 0.759 & 16.53 \\
& $\overline{e^2}$
& 0.887 & 19.21
& 0.849 & 18.44 \\[0.3em]

Directional / advection (100 km, $\pm45^\circ$)
& $\overline e$
& 0.319 & 11.95
& 0.362 & 12.43 \\
& $\overline{e^2}$
& 0.656 & 24.29
& 0.685 & 23.24 \\
\hline
\end{tabular}

\vspace{0.4em}
\parbox{0.97\textwidth}{
\footnotesize
\emph{Notes:}
Moran’s $I$ statistics are reported for station-level mean residuals
($\overline e$) and mean squared residuals ($\overline{e^2}$).
All statistics are computed under randomisation using row-standardised spatial
weight matrices.
Distance-band weights use a 55\,km threshold, $k$-nearest neighbour weights use
$k=5$, and directional (advection-based) weights are constructed using a
100\,km cutoff and a $\pm45^\circ$ upwind cone.
All tests strongly reject the null hypothesis of spatial independence
(p-values $<2.2\times10^{-16}$).
}
\end{table}

Across all weighting schemes and measurement heights, Moran’s $I$ statistics
indicate strong spatial dependence in wind-speed dynamics. Spatial correlation
is consistently stronger for the volatility proxy $\overline{e^2}$ than for the
mean residual $\overline e$, indicating that volatility clustering exhibits a
pronounced spatial component. Dependence is strongest under the
$k$-nearest-neighbour structure, followed by directional and distance-band
specifications, and is slightly more pronounced at the 100\,m measurement height.

\subsection{Estimation and forecasting setup}
To assess the role of spatial dependence in the mean dynamics, the
spatiotemporal volatility model is estimated using two alternative
residual specifications. First, residuals obtained from the station-wise preprocessing procedure described in Section~\ref{sec:univariate_garch} are used as inputs to the volatility model. Second, we consider residuals obtained from a Spatial Dynamic Panel Data (SDPD) mean specification, which explicitly accounts for contemporaneous and lagged spatial dependence in wind-speed levels. This setup also allows us to evaluate how misspecification of the conditional mean may affect the modelling of volatility dynamics. While the STARMAGARCH framework already incorporates spatial dynamics in the mean equation through spatial autoregressive and moving-average terms, the SDPD specification provides a more flexible representation of spatial interactions in the conditional mean. Comparing the volatility model under these two preprocessing strategies, therefore, provides a useful robustness check and allows us to assess whether explicitly modelling spatial dependence in the mean improves the adequacy of the volatility specification and the resulting forecasting performance.
The spatiotemporal mean specification considered in this study follows a
Spatial Dynamic Panel Data (SDPD) \cite[see]{lee2012qml} formulation of the form
\begin{equation}
\mathbf{y}_t =
\rho\,\mathbf{W}\mathbf{y}_t
+ \boldsymbol{\gamma}\odot \mathbf{y}_{t-1}
+ \lambda\,\mathbf{W}\mathbf{y}_{t-1}
+ \boldsymbol{\epsilon}_t ,
\end{equation}
where $\mathbf{W}$ denotes the spatial weight matrix, $ \boldsymbol{\gamma}=(\gamma_1, \cdots, \gamma_N)$, $\rho$, $\lambda$ are real numbers and
$\boldsymbol{\epsilon}_t$ represents the innovation process.
This comparison allows us to assess the extent to which explicitly modelling spatial dependence in the mean equation improves the adequacy of the volatility specification. Estimation is performed using three alternative spatial weight matrices introduced in Section~\ref{sec:spatial_weights_moran}. Parameters are estimated via maximum likelihood using the Template Model Builder (TMB) framework and the authors' implementation  \footnote{\url{https://github.com/holleland/starmagarch}}. For each weight matrix and measurement height, the model is estimated once on the in-sample period 2016--2020. Out-of-sample evaluation relies on a recursive one-step-ahead forecasting scheme with fixed parameters over the test period (2021). Forecasts of the conditional variance are assessed against realised volatility proxies constructed from the residuals $\varepsilon_{i,t}$, including daily squared residuals and an exponentially weighted moving average (EWMA) benchmark defined as \begin{equation} \mathrm{RV}_{i,t} = \varepsilon_{i,t}^2, \qquad \mathrm{EWMA}_{i,t} = \lambda\,\mathrm{EWMA}_{i,t-1} + (1-\lambda)\,\varepsilon_{i,t}^2 , \end{equation} where the smoothing parameter is fixed at $\lambda = 0.94$. Forecast accuracy is evaluated using the root mean squared forecast error (RMSFE) and mean absolute forecast error (MAFE) computed on the logarithmic scale of the conditional variance, \begin{equation} \text{RMSFE} = \sqrt{ \frac{1}{NT} \sum_{i=1}^{N} \sum_{t=1}^{T} \left( \log \hat{h}_t(s_i) - \log \tilde{h}_t(s_i) \right)^2 }, \qquad \text{MAFE} = \frac{1}{NT} \sum_{i=1}^{N} \sum_{t=1}^{T} \left| \log \hat{h}_t(s_i) - \log \tilde{h}_t(s_i) \right|, \label{eq:rmsfemafe} \end{equation} where $T$ denotes the number of forecast periods, $\hat{h}_t(s_i)$ the predicted conditional variance at location $s_i$, and $\tilde{h}_t(s_i)$ a proxy for realised volatility. Table~\ref{tab:main_forecast_results} reports forecasting performance using realised volatility (RV) and EWMA proxies. Results obtained with alternative volatility measures are provided in Appendix~\ref{tab:appendix_forecast_results}. Model adequacy is further assessed using Ljung--Box tests for temporal dependence and Moran's $I$ tests for residual spatial dependence, with diagnostic results reported in Appendix~\ref{tab:appendix_forecast_results}.

\subsection{Estimation and the outcomes of forecasting}

The STARMAGARCH models are estimated using residuals obtained from two
alternative mean specifications: a station-wise AR(1) model and a Spatial
Dynamic Panel Data (SDPD) model. Comparing these two setups allows us to evaluate how the treatment of spatial dependence in the mean equation
affects the estimated volatility dynamics and the resulting forecasting
performance.

Table~\ref{tab:master_comparison} reports parameter estimates for the
STARMAGARCH models across the three spatial weight matrices considered in
this study: distance-based, $k$-nearest neighbour (KNN), and directional
(advection-based) matrices. Results are presented separately for wind
speed measured at 10\,m and 100\,m heights and for both residual
specifications. Table~\ref{tab:diagnostics_comparison_master} summarises
diagnostic tests for temporal and spatial dependence in the residuals and
squared residuals. Finally, Table~\ref{tab:main_forecast_results} reports the out-of-sample forecasting performance based on one-step-ahead variance forecasts evaluated on the logarithmic variance scale. Forecasts are generated using parameters estimated on the training sample (2016--2020) and kept fixed during the evaluation period (2021). 

Across all specifications, the volatility persistence
parameter $\hat{\beta}$ increases systematically with measurement height.
For instance, in the distance-based model estimated on AR(1) residuals,
$\hat{\beta}$ rises from $0.5018$ at 10\,m to $0.7934$ at 100\,m. This indicates
that wind-speed variability at higher altitudes exhibits stronger
temporal persistence, reflecting the more stable atmospheric conditions
away from the surface layer where frictional turbulence dominates. This result is consistent with the well-known structure of the atmospheric boundary layer, where surface friction and obstacles generate strong turbulence close to the ground, while winds at higher altitudes are less affected by surface drag and therefore exhibit smoother and more persistent dynamics \citep{monin1954basic, stull2012introduction}. Consequently, hub-height wind speeds tend to display greater temporal persistence than near-surface winds.
The directional weight matrix produces markedly different behaviour. The persistence parameter $\beta$ is not
statistically significant across all cases, implying that the volatility process
effectively collapses to an ARCH-type structure. This suggests that,
while wind propagation exhibits a directional component, directional
connectivity alone appears insufficient to capture the persistence of
wind-speed variance observed in this dataset.

A central finding concerns the effect of modelling spatial dependence in
the mean equation. When STARMAGARCH is estimated using AR(1) residuals,
strong spatial correlation remains in the residuals. As shown in
Table~\ref{tab:diagnostics_comparison_master}, Moran’s $I$ pass rates are
close to zero in several configurations, indicating that spatial
dependence in wind-speed levels has not been fully removed.

Introducing the SDPD mean specification substantially improves the
diagnostic behaviour of the residuals. By explicitly modelling both
contemporaneous and lagged spatial spillovers, the SDPD model absorbs a
large portion of spatial dependence in the levels. For example, under the
KNN structure Moran’s $I$ pass rates increase from nearly $0\%$ to
$76.84\%$. This suggests that a portion of the volatility clustering
detected in simpler specifications may actually arise from unmodelled
spatial dependence in the mean process rather than genuine variance
dynamics.

The forecasting results reported in
Table~\ref{tab:main_forecast_results} show a clear interaction between
mean specification and the role of spatial structure in volatility modelling.
When STARMAGARCH is estimated using AR(1) residuals, the $k$-nearest-neighbour
specification delivers the best forecasting performance across most proxies and
both measurement heights. This suggests that, in the presence of misspecified
mean dynamics, a flexible local connectivity structure is better able to absorb
remaining cross-sectional dependence.
In contrast, once spatial dependence in the mean equation is explicitly
accounted for through the SDPD specification, the distance-based weight matrix
consistently becomes the most competitive alternative. In particular, the
distance specification combined with SDPD residuals achieves the lowest forecast
errors across all proxies at both heights. This indicates that, once the main
spatial structure in wind-speed levels is properly captured, a parsimonious
distance-based representation is sufficient to describe volatility spillovers.
The directional specification, although motivated by physical considerations,
does not lead to improvements in forecasting performance, suggesting that capturing directional effects may be more relevant for structural
interpretation than for short-term prediction.
These results point to an important modelling insight: the apparent gains from
more flexible spatial structures can largely reflect compensation for a misspecified mean model. When the mean equation is adequately specified, the relative importance of the volatility model shifts towards stability and parsimony rather than flexibility.
Interestingly, the univariate GARCH benchmarks remain highly competitive,
particularly when the EWMA proxy is used as the volatility target. This likely
reflects the strong smoothing inherent in the EWMA construction, which reduces
the impact of spatial interactions and favours simpler time-series-based
volatility models. Additional results based on alternative realised volatility
proxies are reported in Appendix~\ref{tab:appendix_forecast_results}.

\begin{table}[h!]
\centering
\caption{Results of STARMAGARCH models with residuals from AR(1) and SDPD model}
\label{tab:master_comparison}
\footnotesize 
\setlength{\tabcolsep}{3pt} 
\begin{tabular}{ll cc cc @{\hspace{12pt}} cc cc}
\toprule
& & \multicolumn{4}{c}{\textbf{AR residuals}} & \multicolumn{4}{c}{\textbf{SDPD residuals}} \\
\cmidrule(r){3-6} \cmidrule(l){7-10}
& & \multicolumn{2}{c}{\textbf{WS10}} & \multicolumn{2}{c}{\textbf{WS100}} & \multicolumn{2}{c}{\textbf{WS10}} & \multicolumn{2}{c}{\textbf{WS100}} \\
\cmidrule(r){3-4} \cmidrule(lr){5-6} \cmidrule(lr){7-8} \cmidrule(l){9-10}
\textbf{Weight} & \textbf{Param} & \textbf{Est.} & \textbf{P-val} & \textbf{Est.} & \textbf{P-val} & \textbf{Est.} & \textbf{P-val} & \textbf{Est.} & \textbf{P-val} \\
\midrule

\multirow{8}{*}{Distance} 
& $\mu$    & -0.0061 & <0.001 & -0.0227 & <0.001 & 0.0020  & <0.001 & -0.0032 & <0.001 \\
& $\phi$   & -0.5969 & <0.001 & -0.4636 & <0.001 & -0.0945 & 0.118  & 0.0409  & 0.189  \\
& $\theta$ & 0.6601  & <0.001 & 0.5506  & <0.001 & 0.2317  & 0.002  & 0.1862  & <0.001 \\
& $\omega$ & 0.0962  & <0.001 & 0.0998  & <0.001 & 0.0200  & <0.001 & 0.0060  & <0.001 \\
& $\alpha$ & 0.2040  & <0.001 & 0.1289  & <0.001 & 0.2858  & <0.001 & 0.1422  & <0.001 \\
& $\beta$  & 0.5018  & <0.001 & 0.7934  & <0.001 & 0.4278  & <0.001 & 0.8800  & <0.001 \\
\cmidrule(lr){2-10}
& AIC      & \multicolumn{2}{c}{210431.8} & \multicolumn{2}{c}{381664.5} & \multicolumn{2}{c}{-8219.2}  & \multicolumn{2}{c}{99681.8} \\
& BIC      & \multicolumn{2}{c}{210419.8} & \multicolumn{2}{c}{381652.5} & \multicolumn{2}{c}{-8231.2}  & \multicolumn{2}{c}{99669.8} \\

\midrule

\multirow{8}{*}{KNN} 
& $\mu$    & 0.0033  & <0.001 & -0.0190 & <0.001 & 0.0041  & <0.001 & 0.0075  & <0.001 \\
& $\phi$   & -0.6084 & <0.001 & -0.5975 & <0.001 & 0.3001  & <0.001 & 0.4043  & <0.001 \\
& $\theta$ & 0.6557  & <0.001 & 0.6589  & <0.001 & -0.4447 & <0.001 & -0.5411 & <0.001 \\
& $\omega$ & 0.0037  & <0.001 & 0.0252  & <0.001 & 0.0010  & <0.001 & 0.0009  & <0.001 \\
& $\alpha$ & 0.1012  & <0.001 & 0.0796  & <0.001 & 0.2434  & <0.001 & 0.1293  & <0.001 \\
& $\beta$  & 0.8952  & <0.001 & 0.9040  & <0.001 & 0.7369  & <0.001 & 0.8620  & <0.001 \\
\cmidrule(lr){2-10}
& AIC      & \multicolumn{2}{c}{\textbf{201024.6}} & \multicolumn{2}{c}{\textbf{377740.5}} & \multicolumn{2}{c}{\textbf{-161541.8}} & \multicolumn{2}{c}{\textbf{-37587.4}} \\
& BIC      & \multicolumn{2}{c}{\textbf{201012.6}} & \multicolumn{2}{c}{\textbf{377728.5}} & \multicolumn{2}{c}{\textbf{-161553.8}} & \multicolumn{2}{c}{\textbf{-37599.4}} \\

\midrule

\multirow{8}{*}{Directional} 
& $\mu$    & -0.0093 & <0.001 & -0.0299 & <0.001 & 0.0155  & <0.001 & 0.0263  & <0.001 \\
& $\phi$   & -0.6799 & <0.001 & -0.4365 & <0.001 & 0.6572  & <0.001 & 0.5948  & <0.001 \\
& $\theta$ & 0.7358  & <0.001 & 0.5225  & <0.001 & -0.7505 & <0.001 & -0.6837 & <0.001 \\
& $\omega$ & 0.2545  & <0.001 & 0.9979  & <0.001 & 0.1160  & <0.001 & 0.3693  & <0.001 \\
& $\alpha$ & 0.2247  & <0.001 & 0.2215  & <0.001 & 0.0359  & <0.001 & 0.0273  & <0.001 \\
& $\beta$  & 0.0000  & 0.500  & 0.0000  & 0.500  & 0.0000  & 0.500  & 0.0000  & 0.500  \\
\cmidrule(lr){2-10}
& AIC      & \multicolumn{2}{c}{212878.8} & \multicolumn{2}{c}{386637.3} & \multicolumn{2}{c}{91949.9}  & \multicolumn{2}{c}{240063.4} \\
& BIC      & \multicolumn{2}{c}{212866.8} & \multicolumn{2}{c}{386625.3} & \multicolumn{2}{c}{91937.9}  & \multicolumn{2}{c}{240051.4} \\
\bottomrule
\end{tabular}
\end{table}

\begin{table}[h!]
\centering
\caption{Diagnostic Summary: percentage of uncorrelated residuals and squared residuals}
\label{tab:diagnostics_comparison_master}
\small
\setlength{\tabcolsep}{3pt}
\begin{tabular}{ll cc cc @{\hspace{12pt}} cc cc}
\toprule
& & \multicolumn{4}{c}{\textbf{AR(1) Residuals (\% Pass)}} & \multicolumn{4}{c}{\textbf{SDPD Residuals (\% Pass)}} \\
\cmidrule(r){3-6} \cmidrule(l){7-10}
& & \multicolumn{2}{c}{\textbf{WS10}} & \multicolumn{2}{c}{\textbf{WS100}} & \multicolumn{2}{c}{\textbf{WS10}} & \multicolumn{2}{c}{\textbf{WS100}} \\
\cmidrule(r){3-4} \cmidrule(lr){5-6} \cmidrule(lr){7-8} \cmidrule(l){9-10}
\textbf{Weight} & \textbf{Test Type} & \textbf{Res.} & \textbf{Sq.} & \textbf{Res.} & \textbf{Sq.} & \textbf{Res.} & \textbf{Sq.} & \textbf{Res.} & \textbf{Sq.} \\
\midrule
\multirow{2}{*}{Distance} 
& Ljung-Box (T) & 88.65 & 81.56 & 88.65 & 88.65 & 31.21 & 80.85 & 24.11 & 77.30 \\
& Moran's I (S) & 0.05  & 0.27  & 0.00  & 0.11  & 22.66 & 6.33  & 7.76  & 5.67  \\
\midrule
\multirow{2}{*}{KNN} 
& Ljung-Box (T) & 96.45 & 92.91 & 92.20 & 89.36 & 31.91 & 81.56 & 27.66 & 77.30 \\
& Moran's I (S) & 0.00  & 0.00  & 0.00  & 0.05  & 64.69 & 8.75  & 76.84 & 5.28  \\
\midrule
\multirow{2}{*}{Directional} 
& Ljung-Box (T) & 75.18 & 60.99 & 76.60 & 24.82 & 30.50 & 26.24 & 14.18 & 9.22  \\
& Moran's I (S) & 0.44  & 5.04  & 0.60  & 5.91  & 62.71 & 53.91 & 73.27 & 54.57 \\
\bottomrule
\multicolumn{10}{l}{\footnotesize \textit{Note: Values represent the \% of stations passing the test at the 5\% significance level.}} \\
\multicolumn{10}{l}{\footnotesize \textit{Higher percentages indicate better model performance in removing correlations.}} \\
\multicolumn{10}{l}{\footnotesize \textit{(T) denotes Temporal correlation; (S) denotes Spatial correlation.}}
\end{tabular}
\end{table}

\begin{table}[h!]
\centering
\caption{Out-of-sample one-step-ahead forecasting performance measured by RMSFE and MAFE (log variance scale) for wind speed at 10\,m and 100\,m. Forecasts are estimated on the 2016--2020 training sample and evaluated over the 2021 test period.}
\label{tab:main_forecast_results}
\small
\setlength{\tabcolsep}{4pt}

\begin{tabular}{lll cc cc}
\toprule
& & & \multicolumn{2}{c}{\textbf{WS10}} & \multicolumn{2}{c}{\textbf{WS100}} \\
\cmidrule(r){4-5} \cmidrule(l){6-7}
\textbf{Model} & \textbf{Matrix} & \textbf{Proxy} & RMSFE & MAFE & RMSFE & MAFE \\
\midrule

\multicolumn{7}{c}{\textbf{STARMAGARCH (AR(1) residuals)}}\\
\midrule
& Distance & RV   & 3.3087 & 2.4997 & 3.1213 & 2.3122 \\
&          & EWMA & 1.0356 & 0.7639 & 0.7142 & 0.5362 \\
\cmidrule(l){2-7}
& KNN      & RV   & \textbf{3.1109} & \textbf{2.3060} & \textbf{3.0482} & \textbf{2.2410} \\
&          & EWMA & \textbf{0.6464} & \textbf{0.4431} & \textbf{0.5268} & \textbf{0.3721} \\
\cmidrule(l){2-7}
& Directional & RV & 3.3673 & 2.5570 & 3.2275 & 2.4164 \\
&             & EWMA & 1.1142 & 0.8353 & 0.8965 & 0.7053 \\

\midrule
\multicolumn{7}{c}{\textbf{STARMAGARCH (SDPD residuals)}}\\
\midrule
& Distance & RV   & \textbf{3.0107} & \textbf{2.1927} & \textbf{3.0639} & \textbf{2.2470} \\
&          & EWMA & \textbf{0.7198} & \textbf{0.5622} & \textbf{0.6756} & \textbf{0.5299} \\
\cmidrule(l){2-7}
& KNN      & RV   & 3.2356 & 2.4216 & 3.3321 & 2.5130 \\
&          & EWMA & 0.9610 & 0.7766 & 0.9966 & 0.8099 \\
\cmidrule(l){2-7}
& Directional & RV & 3.3072 & 2.4944 & 3.2930 & 2.4989 \\
&             & EWMA & 1.0315 & 0.8120 & 1.0275 & 0.8484 \\

\midrule
\multicolumn{7}{c}{\textbf{Univariate GARCH benchmarks}}\\
\midrule
GARCH & --- & RV   & 3.1009 & 2.3020 & 3.0429 & 2.2382 \\
       & --- & EWMA & \textbf{0.5361} & \textbf{0.4399} & \textbf{0.4603} & \textbf{0.3619} \\
\cmidrule(l){1-7}
EGARCH & --- & RV   & 3.1202 & 2.3203 & 3.0976 & 2.2933 \\
       & --- & EWMA & 0.5904 & 0.4905 & 0.5768 & 0.4708 \\

\bottomrule
\end{tabular}

\vspace{0.3cm}
\parbox{0.95\textwidth}{
\footnotesize
\textit{Notes:} Forecast errors are computed on the logarithmic variance scale.
Bold values indicate the lowest forecasting error within each model block.
}
\end{table}

\section{Multivariate spatiotemporal extension across measurement heights}
\label{sec5:multivariatelogarch}
The preceding analysis treats wind speeds measured at 10,m and 100,m as isolated spatiotemporal systems. While this allows for a direct comparison of volatility dynamics across heights, it does not account for the intrinsic cross-height dependencies. We therefore consider a multivariate framework in which both levels are modeled jointly, accounting for both horizontal interactions across locations and vertical coupling between heights. Formally, we consider the process 
$$\{\mathbf{Y}(\mathbf{s}, t) : \mathbf{s} \in \mathcal{D} \subset \mathbb{R}^d, t \in \mathcal{T} \subset \mathbb{R}\}$$ 
where $\mathbf{Y}$ takes values in $\mathbb{R}^p$. Following the setup in previous sections, let $\tilde{\mathbf{Y}}_t \in \mathbb{R}^{n \times p}$ denote the matrix of deseasonalized wind-speed observations at time $t$, where $n$ is the number of spatial locations and $p=2$ corresponds to the two measurement heights. Building on the multivariate spatiotemporal autoregressive framework of \citet{eckardt2025regional} for the mean and the multivariate spatiotemporal log-ARCH formulation (vec-spARCH) of \citet{otto2024multivariate} for the volatility, we consider the following specification:
\begin{align}
\tilde{\mathbf{Y}}_t
&= \boldsymbol{\beta}_\mu \mathbf{X}_t 
+ \mathbf{W}\tilde{\mathbf{Y}}_t \boldsymbol{\Psi}_{\mu}
+ \tilde{\mathbf{Y}}_{t-1}\boldsymbol{\Pi}_{\mu}
+ \boldsymbol{\varepsilon}_t, \\
\boldsymbol{\varepsilon}_t
&= \mathbf{H}_t^{1/2} \circ \boldsymbol{\Xi}_t, \\
\mathbf{H}_t^{(\ln)}
&= \mathbf{A}
+ \mathbf{W} \boldsymbol{\varepsilon}_{t}^{(\ln,2)} \boldsymbol{\Psi}_{\sigma}
+ \boldsymbol{\varepsilon}_{t-1}^{(\ln,2)} \boldsymbol{\Pi}_{\sigma}.
\end{align}

The spatial weight matrix $\mathbf{W} \in \mathbb{R}^{n \times n}$ captures interactions between monitoring stations. The matrix $\boldsymbol{\Psi}_{\mu} \in \mathbb{R}^{p \times p}$ governs contemporaneous spatial dependence: its diagonal elements describe same-height interactions, while the off-diagonal elements capture cross-height effects transmitted through the spatial network. Temporal dependence is described by $\boldsymbol{\Pi}_{\mu}$, allowing for lagged interactions between heights. The matrix $\mathbf{X}_t \in \mathbb{R}^{n \times p}$ represents exogenous covariates; in this application, only an intercept is included, so that $\boldsymbol{\beta}_{\mu}$ captures height-specific mean levels.
The innovations $\boldsymbol{\varepsilon}_t$ are modelled with time-varying variance $\mathbf{H}_t$, and the variance dynamics are specified in logarithmic form to ensure positivity. The matrix $\boldsymbol{\Psi}_{\sigma}$ captures contemporaneous volatility spillovers, while $\boldsymbol{\Pi}_{\sigma}$ reflects the effect of past squared innovations on current log-volatility. Cross-height transmission is allowed in both components. The disturbances $\boldsymbol{\Xi}_t$ consist of independent and identically distributed (i.i.d.) random vectors. Each vector has a mean of zero ($E[\boldsymbol{\varepsilon}_{j,t}] = \mathbf{0}$) and an identity covariance matrix ($\text{Cov}[\boldsymbol{\varepsilon}_{j,t}] = \mathbf{I}_n$), $j\in {1,\cdots,p}$.
Estimation is performed using the log-squared residual transformation, which yields a linear spatiotemporal representation while preserving the dependence structure. The fitted values correspond to conditional log-volatility, and volatility levels are obtained by exponentiation.

\medskip

The empirical results are reported in Table~\ref{tab:st_multivariate_results}. 
For the distance and $k$-nearest-neighbour matrices, strong contemporaneous spatial dependence is observed within each height, with $\Psi_{\mu,11}$ and $\Psi_{\mu,22}$ taking values close to the upper bound. This indicates a high degree of spatial interdependence across locations at both measurement heights.
Cross-height effects are statistically significant in all cases. For the distance matrix, $\Psi_{\mu,21}$ is substantially larger than $\Psi_{\mu,12}$, indicating a stronger transmission from 100\,m to 10\,m. The same ordering is observed for the $k$-nearest-neighbour specification, although with smaller magnitudes, suggesting that higher-altitude wind dynamics exert a stronger influence on near-surface behaviour than the reverse.
Temporal effects in $\boldsymbol{\Pi}_{\mu}$ are moderate. The diagonal elements indicate persistence within each height, while the off-diagonal terms reveal weaker and sometimes negative cross-height lagged effects, consistent with short-term adjustment dynamics.

In the volatility equation, all elements of $\boldsymbol{\Psi}_{\sigma}$ are positive and highly significant. Volatility shocks propagate across locations and between heights, with comparable magnitudes across all entries. In contrast to the mean equation, the cross-height effects are more balanced, with no clear dominance of one direction. The coefficients of $\boldsymbol{\Pi}_{\sigma}$ are positive but relatively small, indicating limited persistence and supporting a short-memory volatility structure.
For the directional specification, a combined weight matrix is constructed as a weighted average of the directional matrices at 10\,m and 100\,m. The mean dynamics differ more markedly: the cross-height effect $\Psi_{\mu,21}$ becomes strongly negative, while $\Psi_{\mu,12}$ remains positive, highlighting the sensitivity of the mean equation to the choice of spatial structure. In contrast, the volatility parameters remain relatively stable across spatial specifications, both in magnitude and significance.
These results provide clear evidence of cross-height dependence in wind-speed dynamics. The dependence is asymmetric in the mean, with stronger effects from 100\,m to 10\,m, while volatility spillovers are more symmetric across heights.

The multivariate forecasting results, reported in
Appendix~\ref{tab:multivar_forecast_appendix}, show that accounting for
cross-height dependence changes the relative performance of the spatial
weight matrices. In contrast to the univariate STARMAGARCH results, the
combined directional matrix consistently provides lower forecast errors
for most realised volatility proxies. The improvement is particularly
pronounced for the smoothed measures, where the directional specification
performs best at both measurement heights.
For the daily squared residual proxy, the distance matrix remains
competitive at 10\,m, although the directional specification yields lower
absolute errors. At 100\,m, the directional matrix performs best across
all proxies. This suggests that directional information becomes more
relevant when the volatility proxy is less noisy and better reflects the
underlying variability in wind speed.
The diagnostic results in Appendix~\ref{tab:multivar_diag_appendix} are consistent with these findings. While the mean model does not fully remove temporal and spatial
dependence, the multivariate volatility model substantially improves the
residual diagnostics. In particular, Ljung--Box tests applied to squared
residuals show high pass rates, and Moran’s $I$ statistics indicate that
most of the spatial dependence in volatility has been removed. This
suggests that the model captures the main features of the joint
spatiotemporal variability.

Overall, these results indicate that the role of spatial structure depends
on how dependence is represented in the model. When interactions between
measurement heights are explicitly included, directional effects appear
more informative for prediction. In contrast, in the univariate setting,
simpler distance-based structures provide a more robust representation of
volatility spillovers.

\begin{table}[htbp]
\centering
\caption{Estimated parameters of the multivariate spatiotemporal model with their standard errors in parentheses.}
\label{tab:st_multivariate_results}
\small
\setlength{\tabcolsep}{6pt}
\begin{tabular}{lccc}
\toprule
\textbf{Parameter} & \textbf{Distance} & \textbf{kNN} & \textbf{Directional} \\
\midrule
\multicolumn{4}{c}{\textit{Mean equation}} \\
\midrule
\textit{Intercept terms} \\

$\beta_{\mu,1}$ & -0.134 (0.002) & 0.007 (0.001) & 0.172 (0.003) \\
$\beta_{\mu,2}$ & -0.333 (0.002) & -0.063 (0.001) & 0.310 (0.004) \\

\addlinespace
\textit{Same-height spatial effects} \\
$\Psi_{\mu,11}$ & 0.906 (0.004) & 0.798 (0.002) & 0.374 (0.005) \\
$\Psi_{\mu,22}$ & 0.848 (0.002) & 0.890 (0.001) & 0.932 (0.003) \\

\addlinespace
\textit{Cross-height spatial effects} \\
$\Psi_{\mu,12}$ & \textbf{0.011 (0.002)} & \textbf{0.072 (0.001)} & \textbf{0.159 (0.003)} \\
$\Psi_{\mu,21}$ & \textbf{0.307 (0.004)} & \textbf{0.166 (0.003)} & \textbf{-0.480 (0.006)} \\

\addlinespace
\textit{Temporal effects} \\
$\Pi_{\mu,11}$ & 0.288 (0.003) & 0.152 (0.002) & 0.338 (0.005) \\
$\Pi_{\mu,22}$ & 0.035 (0.001) & 0.051 (0.001) & 0.147 (0.002) \\
$\Pi_{\mu,12}$ & -0.079 (0.001) & -0.047 (0.001) & -0.049 (0.002) \\
$\Pi_{\mu,21}$ & 0.101 (0.003) & -0.022 (0.002) & 0.162 (0.005) \\

\midrule
\multicolumn{4}{c}{\textit{Volatility equation}} \\
\midrule
\textit{Intercepts} \\
$A_{1}$ & -0.835 (0.017) & -1.432 (0.022) & -1.072 (0.011) \\
$A_{2}$ & -0.372 (0.019) & -0.859 (0.026) & -0.325 (0.012) \\

\addlinespace
\textit{Same-height spillovers} \\
$\Psi_{\sigma,11}$ & 0.558 (0.004) & 0.466 (0.003) & 0.535 (0.004) \\
$\Psi_{\sigma,22}$ & 0.577 (0.004) & 0.454 (0.003) & 0.519 (0.004) \\

\addlinespace
\textit{Cross-height spillovers} \\
$\Psi_{\sigma,12}$ & \textbf{0.163 (0.005)} & \textbf{0.122 (0.006)} & \textbf{0.108 (0.004)} \\
$\Psi_{\sigma,21}$ & \textbf{0.135 (0.006)} & \textbf{0.116 (0.006)} & \textbf{0.166 (0.004)} \\

\addlinespace
\textit{Lagged shock effects} \\
$\Pi_{\sigma,11}$ & 0.067 (0.002) & 0.107 (0.002) & 0.113 (0.002) \\
$\Pi_{\sigma,22}$ & 0.113 (0.002) & 0.171 (0.002) & 0.085 (0.002) \\
$\Pi_{\sigma,12}$ & 0.054 (0.002) & 0.066 (0.002) & 0.050 (0.002) \\
$\Pi_{\sigma,21}$ & 0.022 (0.002) & 0.053 (0.002) & 0.065 (0.002) \\

\bottomrule
\end{tabular}
\end{table}

\section{Conclusion}
\label{sec:conclusion}

This paper investigated the temporal and spatial dynamics of wind-speed volatility across a dense monitoring network in Northern Italy, using daily observations from 141 stations at 10\,m and 100\,m measurement heights. The analysis focused on how spatial interactions and vertical dependence influence the modelling and forecasting of wind-speed variability.
The univariate results confirmed the presence of pronounced volatility clustering, but also showed that substantial spatial dependence remains in residuals when mean dynamics are modelled independently across locations. Incorporating a Spatial Dynamic Panel Data (SDPD) specification in the mean equation significantly improves residual diagnostics, indicating that a large share of the apparent volatility clustering is attributable to unmodelled spatial dependence in wind-speed levels.
Within the STARMAGARCH framework, the results highlight that the role of the spatial weight matrix depends critically on the specification of the mean equation. When residual spatial dependence persists, more flexible structures such as the $k$-nearest-neighbour matrix provide better forecasting performance. However, once spatial interactions are adequately captured in the mean, a parsimonious distance-based matrix consistently yields the most reliable out-of-sample forecasts. This suggests that, conditional on a well-specified mean, stable and interpretable spatial structures are sufficient to capture volatility spillovers. More broadly, these findings provide a unified perspective linking mean specification, spatial structure, and volatility forecasting performance in spatiotemporal wind-speed models.

The analysis also reveals systematic differences across measurement heights. Volatility persistence increases with height, indicating more stable and predictable dynamics at turbine hub levels, while near-surface wind speeds are more affected by local turbulence and short-term fluctuations. The multivariate extension further provides evidence of cross-height dependence, with asymmetric interactions in the mean, dominated by effects from 100\,m to 10\,m, and more balanced spillovers in the volatility dynamics.
From an applied perspective, these findings have direct implications for wind energy systems and electricity markets. Improved modelling of wind-speed volatility enhances short-term forecasting accuracy, which is essential for power system balancing and operational planning. The higher persistence observed at turbine hub height suggests that variability at operational levels is more predictable, supporting more reliable production and risk management decisions. Moreover, the presence of spatial and cross-height dependence indicates that wind variability should be treated as a networked process rather than as a collection of independent time series.
Future research could extend this framework by considering hybrid spatial structures that combine distance-based proximity with directional information, as well as by incorporating additional meteorological covariates. An important direction is the development of fully multivariate spatiotemporal GARCH models capable of jointly capturing persistence and asymmetric spillovers across both space and measurement heights.

\section*{Acknowledgement}
We gratefully acknowledge funding by the Deutsche Forschungsgemeinschaft (DFG, German Research Foundation) project number 501539976.

\newpage

\bibliography{reference}

\newpage
\begin{appendix}

\section{Additional results and material}

The full set of scripts used for data preprocessing and for reproducing the empirical results presented in this work is publicly available at:
\begin{center}
\url{https://github.com/ariane237/Spatiotemporal-dynamics-of-wind-speed-volatility}
\end{center}

\medskip

\begin{figure}[htbp]
\centering

\begin{subfigure}[b]{0.35\textwidth}
  \centering
  \includegraphics[width=\textwidth]{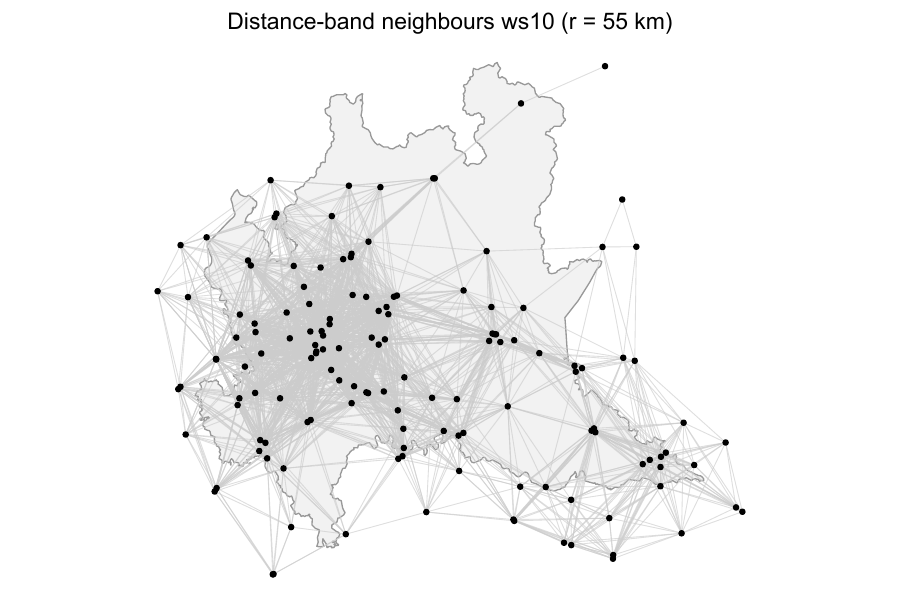}
\end{subfigure}
\hspace{-1.1cm}
\begin{subfigure}[b]{0.35\textwidth}
  \centering
  \includegraphics[width=\textwidth]{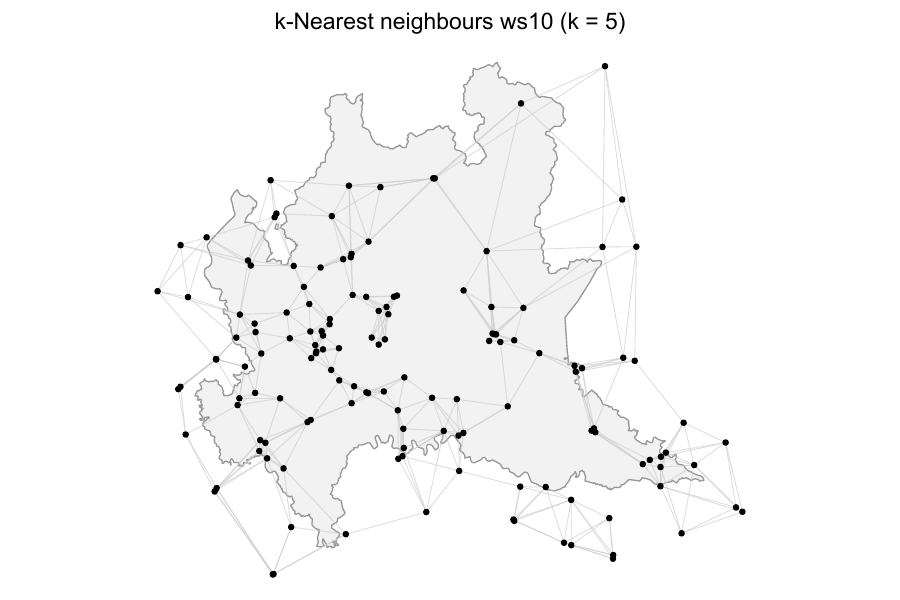}
\end{subfigure}
\hspace{-1.1cm}
\begin{subfigure}[b]{0.35\textwidth}
  \centering
  \includegraphics[width=\textwidth]{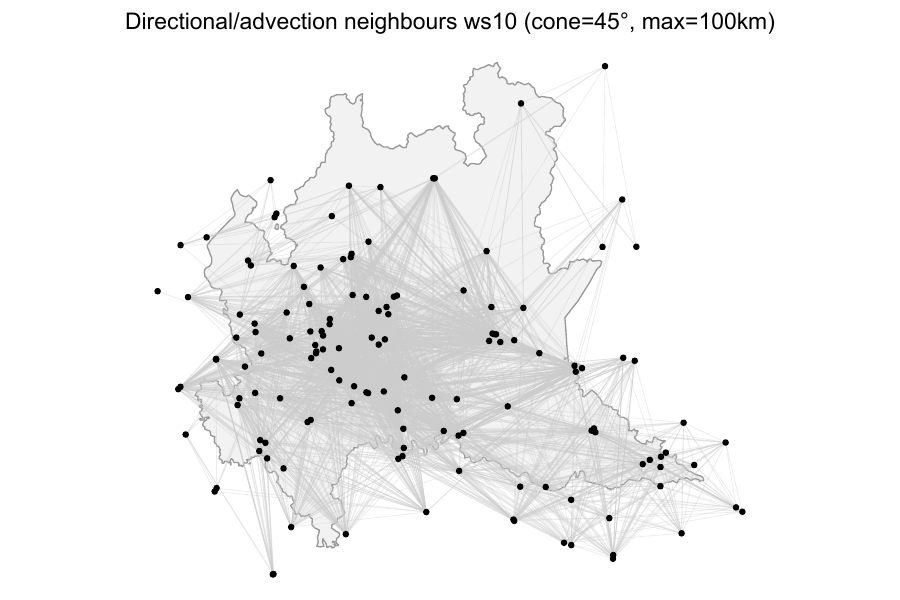}
\end{subfigure}
\hspace{-1.1cm}
\begin{subfigure}[b]{0.35\textwidth}
  \centering
  \includegraphics[width=\textwidth]{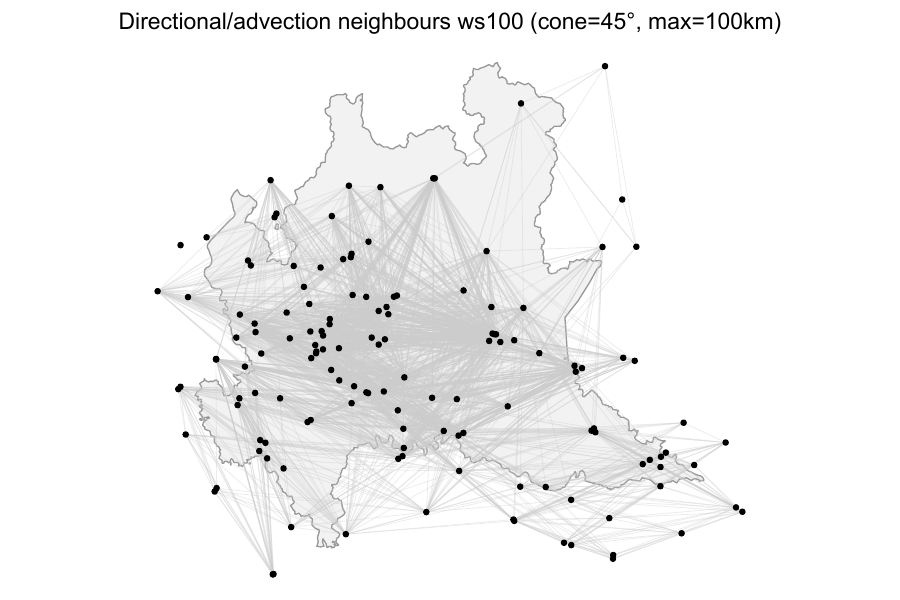}
\end{subfigure}
\caption{Spatial interaction networks implied by the alternative spatial weight matrices for the Lombardy wind monitoring stations.}
\label{fig:W_networks_row}
\end{figure}

Figure~\ref{fig:W_networks_row} illustrates the spatial interaction networks implied by the alternative spatial weight matrices considered in the analysis. Distance-band and $k$-nearest-neighbour constructions reflect purely geometric proximity between monitoring stations, whereas the directional (advection-based) structure incorporates prevailing wind directions together with distance decay, leading to height-specific interaction patterns.

\medskip

Additional forecasting results are reported using alternative realised volatility proxies constructed from the residuals $\varepsilon_{i,t}$. In particular, we consider smoothed measures based on five-day rolling averages of squared and absolute residuals. The first corresponds to a low-frequency realised variance proxy, while the second provides a more robust measure based on absolute returns, which has been shown to be less sensitive to extreme observations \citep{barndorff2002econometric, taylor2011asset}. These are defined as
\begin{equation}
\mathrm{RV5}^{\mathrm{sq}}_{i,t} = \frac{1}{5}\sum_{j=0}^{4} \varepsilon_{i,t-j}^2,
\quad
\mathrm{RV5}^{\mathrm{abs}}_{i,t} = \left( \frac{1}{5}\sum_{j=0}^{4} \lvert \varepsilon_{i,t-j} \rvert \right)^2.
\end{equation}

\medskip

\begin{table}[h!]
\centering
\caption{One-step-ahead forecasting performance using alternative realised volatility proxies.}
\label{tab:appendix_forecast_results}
\small
\setlength{\tabcolsep}{4pt}
\begin{tabular}{lll cc cc}
\toprule
& & & \multicolumn{2}{c}{\textbf{WS10}} & \multicolumn{2}{c}{\textbf{WS100}} \\
\cmidrule(r){4-5} \cmidrule(l){6-7}
\textbf{Model} & \textbf{Matrix} & \textbf{Proxy} & RMSFE & MAFE & RMSFE & MAFE \\
\midrule

\multirow{6}{*}{\shortstack[l]{STARMAGARCH \\ (AR(1) residuals)}} 
& Distance & $\mathrm{RV5}^{\mathrm{abs}}$ & 1.802 & 1.468 & 1.523 & 1.245 \\
&          & $\mathrm{RV5}^{\mathrm{sq}}$  & 1.519 & 1.187 & 1.247 & 0.980 \\
\cmidrule(l){2-7}
& KNN      & $\mathrm{RV5}^{\mathrm{abs}}$ & \textbf{1.498} & \textbf{1.213} & \textbf{1.419} & \textbf{1.155} \\
&          & $\mathrm{RV5}^{\mathrm{sq}}$  & \textbf{1.227} & \textbf{0.954} & \textbf{1.151} & \textbf{0.904} \\
\cmidrule(l){2-7}
& Directional & $\mathrm{RV5}^{\mathrm{abs}}$ & 1.912 & 1.573 & 1.720 & 1.431 \\
&             & $\mathrm{RV5}^{\mathrm{sq}}$  & 1.632 & 1.297 & 1.443 & 1.167 \\

\midrule

\multirow{6}{*}{\shortstack[l]{STARMAGARCH \\ (SDPD residuals)}} 
& Distance & $\mathrm{RV5}^{\mathrm{abs}}$ & \textbf{1.445} & \textbf{1.177} & \textbf{1.475} & \textbf{1.201} \\
&          & $\mathrm{RV5}^{\mathrm{sq}}$  & \textbf{1.194} & \textbf{0.942} & \textbf{1.226} & \textbf{0.972} \\
\cmidrule(l){2-7}
& KNN      & $\mathrm{RV5}^{\mathrm{abs}}$ & 1.705 & 1.409 & 1.816 & 1.499 \\
&          & $\mathrm{RV5}^{\mathrm{sq}}$  & 1.428 & 1.141 & 1.544 & 1.241 \\
\cmidrule(l){2-7}
& Directional & $\mathrm{RV5}^{\mathrm{abs}}$ & 1.843 & 1.546 & 1.863 & 1.587 \\
&             & $\mathrm{RV5}^{\mathrm{sq}}$  & 1.567 & 1.274 & 1.587 & 1.318 \\

\midrule

GARCH  & --- & $\mathrm{RV5}^{\mathrm{abs}}$ & 1.498 & 1.265 & 1.417 & 1.183 \\
       &     & $\mathrm{RV5}^{\mathrm{sq}}$  & \textbf{1.216} & \textbf{0.994} & \textbf{1.146} & \textbf{0.934} \\
\midrule
EGARCH & --- & $\mathrm{RV5}^{\mathrm{abs}}$ & 1.518 & 1.289 & 1.480 & 1.250 \\
       &     & $\mathrm{RV5}^{\mathrm{sq}}$  & 1.231 & 1.009 & 1.197 & 0.982 \\

\bottomrule
\end{tabular}
\end{table}
\medskip

\begin{table}[htbp]
\centering
\caption{Diagnostic summary for the multivariate spatiotemporal model. Values represent the percentage of series (Ljung--Box) or time points (Moran's $I$) passing the test at the 5\% significance level.}
\label{tab:multivar_diag_appendix}
\small
\setlength{\tabcolsep}{4pt}
\begin{tabular}{lll cc cc}
\toprule
& & & \multicolumn{2}{c}{\textbf{WS10}} & \multicolumn{2}{c}{\textbf{WS100}} \\
\cmidrule(r){4-5} \cmidrule(l){6-7}
\textbf{Stage} & \textbf{Weight} & \textbf{Test} & Res. & Sq. & Res. & Sq. \\
\midrule

\multirow{6}{*}{Mean model}
& \multirow{2}{*}{Distance}
& Ljung--Box & 16.3 & 14.2 & 22.7 & 24.8 \\
& & Moran's $I$ & 13.5 & 7.4 & 6.7 & 7.7 \\

\cmidrule(l){2-7}
& \multirow{2}{*}{KNN}
& Ljung--Box & 17.0 & 32.6 & 24.8 & 40.4 \\
& & Moran's $I$ & 46.5 & 2.0 & 87.0 & 1.4 \\

\cmidrule(l){2-7}
& \multirow{2}{*}{Directional}
& Ljung--Box & 7.8 & 10.6 & 12.8 & 14.9 \\
& & Moran's $I$ & 38.7 & 62.3 & 50.9 & 64.0 \\

\midrule

\multirow{6}{*}{Volatility model}
& \multirow{2}{*}{Distance}
& Ljung--Box & 61.7 & 90.8 & 37.6 & 90.1 \\
& & Moran's $I$ & 95.6 & 67.5 & 93.5 & 63.8 \\

\cmidrule(l){2-7}
& \multirow{2}{*}{KNN}
& Ljung--Box & 19.9 & 90.8 & 4.3 & 83.7 \\
& & Moran's $I$ & 95.6 & 79.7 & 96.3 & 81.4 \\

\cmidrule(l){2-7}
& \multirow{2}{*}{Directional}
& Ljung--Box & 26.2 & 94.3 & 41.1 & 87.9 \\
& & Moran's $I$ & 87.2 & 75.7 & 80.0 & 67.7 \\

\bottomrule
\end{tabular}
\end{table}

\medskip

\begin{table}[htbp]
\centering
\caption{Out-of-sample one-step-ahead forecasting performance of the multivariate spatiotemporal volatility model.}
\label{tab:multivar_forecast_appendix}
\small
\setlength{\tabcolsep}{4pt}
\begin{tabular}{ll cc cc}
\toprule
& & \multicolumn{2}{c}{\textbf{WS10}} & \multicolumn{2}{c}{\textbf{WS100}} \\
\cmidrule(r){3-4} \cmidrule(l){5-6}
\textbf{Weight} & \textbf{Proxy} & RMSFE & MAFE & RMSFE & MAFE \\
\midrule

\multirow{4}{*}{Distance}
& RV        & 2.330 & 1.722 & 2.408 & 1.929 \\
& $\mathrm{RV5}^{\mathrm{sq}}$   & 1.322 & 1.049 & 1.958 & 1.732 \\
& $\mathrm{RV5}^{\mathrm{abs}}$  & 1.077 & 0.840 & 1.637 & 1.405 \\
& EWMA      & 1.460 & 1.241 & 2.170 & 2.012 \\

\midrule

\multirow{4}{*}{KNN}
& RV        & 2.406 & 1.822 & 2.521 & 1.970 \\
& $\mathrm{RV5}^{\mathrm{sq}}$   & 1.588 & 1.303 & 1.964 & 1.669 \\
& $\mathrm{RV5}^{\mathrm{abs}}$  & 1.276 & 1.020 & 1.605 & 1.316 \\
& EWMA      & 1.814 & 1.601 & 2.230 & 2.011 \\

\midrule

\multirow{4}{*}{\textbf{Directional}}
& RV        & 2.417 & 1.708 & \textbf{2.302} & \textbf{1.650} \\
& $\mathrm{RV5}^{\mathrm{sq}}$   & \textbf{1.003} & \textbf{0.785} & \textbf{1.072} & \textbf{0.847} \\
& $\mathrm{RV5}^{\mathrm{abs}}$  & \textbf{0.894} & \textbf{0.700} & \textbf{0.881} & \textbf{0.692} \\
& EWMA      & \textbf{1.021} & \textbf{0.828} & \textbf{1.145} & \textbf{0.970} \\

\bottomrule
\end{tabular}

\end{table}

\end{appendix}

\end{document}